\documentclass[final,3p,times]{elsarticle}
\usepackage[T1]{fontenc}
\usepackage{amsfonts}
\usepackage{amssymb}
\usepackage{amsmath}
\usepackage{amsthm}
\usepackage{graphics}
\usepackage{graphicx}
\usepackage{xcolor}
\usepackage{braket, bm}
\usepackage[colorlinks = true, linkcolor = blue, urlcolor = blue, citecolor = blue]{hyperref}
\usepackage{tikz}
\usetikzlibrary{arrows,snakes,backgrounds}
\usepackage{tikzpeople}
\usetikzlibrary{shapes.callouts}
\usepackage{marvosym}
\usepackage{wasysym}
\usepackage{subfigure}

\makeatletter
\def\ps@pprintTitle{   
\let\@oddhead\@empty   
\let\@evenhead\@empty   
\let\@oddfoot\@empty   
\let\@evenfoot\@oddfoot
}
\makeatother

\begin{document}
\begin{frontmatter}
\title{Optimal teleportation fidelity and its deviation in noisy scenarios}

\author{Pratapaditya Bej}
\ead{pratap6906@gmail.com}
\address{Department of Physics and Centre for Astroparticle Physics and Space Science, Bose Institute, EN-80, Sector V,\\  Bidhannagar, Kolkata 700091, India}

\author{Saronath Halder}
\ead{saronath.halder@gmail.com}
\address{Quantum Information and Computation Group, Harish-Chandra Research Institute, HBNI, Chhatnag Road, Jhunsi,\\ Prayagraj (Allahabad) 211019, India}

\author{Ritabrata Sengupta}
\ead{rb@iiserbpr.ac.in}
\address{Department of Mathematical Sciences, Indian Institute of Science Education and Research Berhampur, Transit Campus,\\ Government ITI, Berhampur 760010, Odisha, India}

\begin{abstract}
In this work, we study the combined effects of noisy resource state and noisy classical communication on teleportation fidelity and its deviation. Basically, we consider a teleportation protocol, where a general two-qubit state in canonical form is used as resource, which of course, can be a noisy entangled state. Thereafter, to teleport an unknown qubit, Alice measures her qubits in Bell basis and convey the measurement outcome to Bob via noisy classical channel(s). In particular, we derive the exact formulae of optimal teleportation fidelity and corresponding fidelity deviation where the resource state and the classical communication, both of them can be noisy. We further find conditions for non-classical fidelity and dispersion-free teleportation within the present protocol. In this way, we identify the noisy environments where it is possible to achieve the dispersion-free teleportation without compromising the non-classical fidelity. We also exhibit scenarios where the increase of entanglement in the resource state, may degrade the quality of teleportation. Finally, we discuss on minimum classical communication cost required to achieve non-classical fidelity in our protocol.
\end{abstract}

\end{frontmatter}

\section{Introduction}\label{sec1}
Quantum teleportation \cite{Bennett93, Bouwmeester97, Boschi98, Braunstein98, Dur00, Ma12, Nolleke13} is a physical process via which it is possible to transfer the state of a quantum system from one location to another without knowing the state. To accomplish this process, one can use entanglement \cite{Horodecki09-1, Guhne09, Das17} as resource along with local operations and classical communication (LOCC) \cite{Chitambar14}. This process is important for many reasons. Two of the reasons are (i) state can be transferred from one location to another without transmitting the quantum system, (ii) any nonlocal task can be accomplished by a teleportation-based protocol \cite{Chefles00-1, Collins01}. Here a nonlocal task means, the task which cannot be completed by LOCC only. 

For perfect teleportation of an unknown qubit, the spatially separated parties can use a two-qubit maximally entangled state. However, if two resource states are there then it is important to find which resource state is the better one to teleport an unknown qubit. For this purpose, one can use the standard figure of merit, the teleportation fidelity \cite{Horodecki96, Horodecki99}. We usually take the average value of the teleportation fidelity, known as the average fidelity and this average is taken over all input states, uniformly distributed in the Bloch sphere. We mention here that a different figure of merit is introduced in Ref.~\cite{Cavalcanti17}, known as the random teleportation robustness. But here we stick to teleportation fidelity because the fluctuations, associated with this figure of merit, are easier to realize.

Recently, fidelity deviation in quantum teleportation has got considerable attention \cite{Bang12, Bang18, Ghosal20-1, Ghosal20, Roy20, Roy20-1, Ghosal21, Song21}. It is important to study this quantity as it gives the idea of fluctuations associated with teleportation fidelity. In particular, due to unavoidable imperfections which occur in a practical situation, it is expected that fluctuations must be associated with teleportation fidelity. Thus, the duo, teleportation fidelity and its deviation contain more information to characterize quantum teleportation than the teleportation fidelity alone. In the articles, just cited, it has also been argued that the states which can lead to both non-classical fidelity and zero fidelity deviation, are the most desired states as resource in quantum teleportation. We mention that the condition for zero fidelity deviation is also known as the universality condition which leads to dispersion-free teleportation. 

One can get zero fidelity deviation without compromising the average teleportation fidelity by using twirling operations. But one has to use extra amount of classical communication (in this regard, see also the discussion, given in Ref.~\cite{Ghosal20-1}). This extra amount of classical communication is necessary to disclose the identity of the unitary operation in each step. So, when the classical communication is limited (particularly, not more than two classical bits of information, communicated from Alice's side to Bob's side), it is important to study the effect of noisy classical communication in the teleportation scenario. Again, it is yet to explore how the noisy resource state and the noisy classical communication together affect the average value of the teleportation fidelity and the fidelity deviation in a teleportation protocol. So, we derive the exact formulae of optimal teleportation fidelity and corresponding fidelity deviation for a general two-qubit state as resource when the resource state and the classical communication, both of them can be noisy. 

In a perfect teleportation process, while teleporting an unknown qubit, along with a two-qubit maximally entangled state, shared between Alice and Bob, two classical bits are also required to communicate Alice's measurement outcome to Bob. But in a practical situation, one cannot deny the presence of noise, associated with the resource state as well as with the classical channels. There are a few papers \cite{Banik13, Weinar13}, where noisy classical communication is considered in a teleportation scenario and corresponding fidelity has been calculated but there the resource states are of particular types. Thus, to realize the combined effects of any noisy resource state along with the noisy classical communication on teleportation fidelity and its deviation, it is important to consider a general two-qubit state as resource along with the noisy classical communication. 

In the present teleportation protocol, the resource state is a general two-qubit state but it is in the canonical form. First, Alice measures her qubits in the Bell basis and then, sends the measurement outcome (two classical bits of information) to Bob through noisy classical channel(s). After receiving the message, Bob applies a Pauli operator. Using this protocol, we derive the exact formulae of teleportation fidelity and its deviation. We also prove that the present teleportation fidelity is an optimal one, i.e., given the conditions (two cbits of communication from Alice's side to Bob's side through noisy classical channel(s) of the present kind), the present teleportation fidelity is the maximum achievable fidelity. We further find the conditions for non-classical fidelity and zero fidelity deviation within our protocol. We identify the noisy environments where it is possible to achieve the dispersion-free teleportation without compromising the non-classical fidelity. Interestingly, we exhibit scenarios within the present protocol, where the fidelity deviation increases if the entanglement of the resource state is increased. Thus, the quality of teleportation may degrade in those cases. We also add some discussions on minimum classical communication cost required to achieve non-classical fidelity in the present protocol.

We organize the paper in the following way: In Sec.~\ref{sec2}, a few preliminary concepts are provided. Thereafter, in Sec.~\ref{sec3}, we present the main results with corresponding discussions. Finally, in Sec.~\ref{sec4}, the conclusion is drawn, mentioning some open problems.

\section{Preliminaries}\label{sec2}
Starting from an arbitrary two-qubit resource state $\rho$, if the parties (Alice and Bob, between which the resource state is shared), follow the standard teleportation protocol \cite{Bennett93}, then it may not always be possible to achieve the optimal teleportation fidelity \cite{Horodecki96}. But it is always possible to maximize the average fidelity, if the parties start the protocol with a resource state which is in the canonical form, $\varrho$. Thus, the maximum value of the average fidelity, $\mathbf{F}_{\rho}$ is an optimized value and it is known as the maximal fidelity. This maximum value can be achieved through the protocol of \cite{Bennett93} if the resource state is in the canonical form, as proposed in Ref.~\cite{Badziag00}. So, the quantity of interest is the maximum value of $\mathbf{F}_\rho=\mathbf{F}_\varrho$ (for a greater details, one can go through the paper \cite{Ghosal20-1}, and the references therein). Note that $\mathbf{F}_\varrho$ is simply the average fidelity corresponding to the state $\varrho$. 

In this work, we stick to a protocol which is basically the standard teleportation protocol as given in \cite{Bennett93}. But here the resource state is an arbitrary two-qubit state and it is given in the canonical form. Again, after measurement, Alice sends her measurement outcome to Bob via noisy classical channel(s). However, we now move forward for a greater mathematical details. 

An arbitrary two-qubit state $\rho$ can be expressed in the following form, known as the Hilbert-Schmidt representation:
\begin{equation} \label{eq1}
\rho = \frac{1}{4}\left(\mathbb{I}_4 + \mathbf{R}\boldsymbol{\cdot\sigma}\otimes\mathbb{I}_2 + \mathbb{I}_2\otimes\mathbf{S}\boldsymbol{\cdot\sigma} + \sum_{i,j=1}^3 \mathbb{T}_{ij}\sigma_{i}\otimes\sigma_{j}\right),
\end{equation}
where $\mathbb{I}_n$ is an $n\times n$ identity matrix (in the whole manuscript), \textbf{R}, \textbf{S} are the vectors of $\mathbb{R}^3$, $\mathbf{R(S)}\boldsymbol{\cdot\sigma} = \sum_{i = 1}^3\mbox{R}_i(\mbox{S}_i)\sigma_i$, the elements $\mathbb{T}_{ij} = \mbox{Tr}(\rho\sigma_i\otimes\sigma_j)$, $\forall i,j= 1,2,3$, build a $3\times3$ matrix $\mathbb{T}$. This matrix is real and it is known as the correlation matrix. In the above expression $\sigma_1$, $\sigma_2$, $\sigma_3$ are the standard Pauli matrices \cite{Horodecki96, Badziag00}. We mention here that $\mathbb{I}_2$ is sometimes considered as the zeroth Pauli matrix, $\sigma_0$. Let $\mathbf{t}_{11}, \mathbf{t}_{22}, \mathbf{t}_{33}$ be the eigenvalues of the correlation matrix $\mathbb{T}$. Then the canonical form $\varrho$, of the state $\rho$, is given by-
\begin{equation} \label{eq2}
\varrho = \frac{1}{4}\left(\mathbb{I}_4 + \mathbf{r}\boldsymbol{\cdot\sigma}\otimes\mathbb{I}_2 + \mathbb{I}_2\otimes\mathbf{s}\boldsymbol{\cdot\sigma} + \sum_{i=1}^3\lambda_i|\mathbf{t}_{ii}|\sigma_{i}\otimes\sigma_{i}\right),
\end{equation}
in the above $\lambda_i$ can be $\pm1$. These values can be determined by the sign of the determinant of $\mathbb{T}$, i.e., $\mbox{det}\mathbb{T}$. If $\mbox{det}\mathbb{T}\leq0$, then we can take $\lambda_i = -1$ for all $|\mathbf{t}_{ii}|\neq0$, $i = 1,2,3$. Again, if $\mbox{det}\mathbb{T}>0$, then we can take $\lambda_i,\lambda_j=-1$, and $\lambda_k=+1$ for the choices where $i\neq j\neq k$, $i,j,k = 1,2,3$, while the eigenvalues of $\mathbb{T}$ obey the condition $|\mathbf{t}_{ii}|\geq|\mathbf{t}_{jj}|\geq|\mathbf{t}_{kk}|$ (also see \cite{Ghosal20-1} in this regard). 

Using the state $\varrho$, as resource, the standard teleportation protocol begins. The very first step of which is a measurement in the Bell basis by Alice. To convey the measurement outcome, Alice sends two classical-bits (in other words, cbits) to Bob. But here the classical communication is noisy. However, based on the classical message that Bob receives, he tries to apply the appropriate Pauli's unitary operator to his qubit. In the Table \ref{tab1}, we provide the Bell states, the cbits which is to be communicated by Alice to Bob, and the Pauli's unitary operators. We mention here that for the simplicity of the calculation, we use the notation $\varsigma_k$ for the unitary operators. They are nothing but Pauli's matrices. However, to achieve optimality of teleportation fidelity based on the type of noisy classical communication is given, it might be necessary to change from one strategy of unitary corrections to another. In Sec.~\ref{sec3}, a discussion related to this is given.
\begin{table}[h!]
\begin{center}
\begin{tabular}{|c|c|c|}
\hline
The Bell states & cbits & Unitary operators\\[0.5 ex]
\hline
$|\phi_0\rangle = \frac{1}{\sqrt{2}}\left(\ket{00}+\ket{11}\right)$ & 00 & $\varsigma_0$\\[1 ex]\hline
$|\phi_1\rangle = \frac{1}{\sqrt{2}}(\ket{00}-\ket{11})$ & 11 & $\varsigma_1$\\[1 ex]\hline
$|\phi_2\rangle = \frac{1}{\sqrt{2}}(\ket{01}+\ket{10})$ & 01 & $\varsigma_2$\\[1 ex]\hline
$|\phi_3\rangle = \frac{1}{\sqrt{2}}(\ket{01}-\ket{10})$ & 10 & $\varsigma_3$\\[1 ex] 
\hline
\end{tabular}
\caption{The Bell state $\ket{\phi_k}$ corresponds to the measurement outcome $k$, $\forall k = 0,\dots,3$. Corresponding to each outcome $k$, cbits and the unitary operators are also given.}\label{tab1}
\end{center}
\end{table}



\begin{figure}[h]
\centering
\begin{minipage}{.5\textwidth}
  \centering
\begin{tikzpicture}
\draw[snake=snake,red,ultra thick] (0,0) -- (3.5,0);
\shade[ball color=blue] (3.5,0) circle (.2cm);
\shade[ball color=magenta] (0,0) circle (.2cm);
\node[name=alice, shape=alice, minimum size=0.7cm, hair=purple, shirt=magenta,skin=pink, label=below:Alice]    at (-1,0.0) (10pt) {};
\node[bob,undershirt=black, mirrored, shirt=blue,minimum size=.7cm,label=below:Bob]   at (4.3,0) {};
\node[thick,black,label=below:Quantum state ($\varrho$)]   at (1.8,-0.2) {};
\draw[-{Implies},double,ultra thick,cyan] (1.8,-1)-- (1.8,-1.5);
\node[ultra thick,black,label=above:\textbf{(a).}]   at (-2.4,-0.5) {};
\end{tikzpicture}

\begin{tikzpicture}
\draw[snake=snake,red,ultra thick] (0,0) -- (3.5,0);
\shade[ball color=blue] (3.5,0) circle (.2cm);
\shade[ball color=magenta] (0,0) circle (.2cm);
\shade[ball color=orange] (-0.6,0) circle (.2cm);
\node[name=alice, shape=alice, minimum size=0.7cm, hair=purple, shirt=magenta,skin=pink, label=below:Alice]    at (-1.2,0.0) (10pt) {};
\node[bob,undershirt=black, mirrored, shirt=blue,minimum size=.7cm,label=below:Bob]   at (4.3,0) {};
\node[thick,black,label=below:Quantum state ($\varrho$)]   at (1.8,-0.2) {};
\draw[-{Implies},double,ultra thick,cyan] (1.8,-1)-- (1.8,-1.5);
\node[thick,black,label=above: $|\psi\rangle$]   at (-0.6,0.05) {};
\node[ultra thick,black,label=above:\textbf{(b).}]   at (-2.4,-0.5) {};
\end{tikzpicture}

\begin{tikzpicture}
\draw[snake=snake,red,ultra thick] (0,0) -- (3.5,0);
\shade[ball color=blue] (3.5,0) circle (.2cm);
\shade[ball color=magenta] (0,0) circle (.2cm);
\shade[ball color=orange] (-0.6,0) circle (.2cm);
\node[name=alice, shape=alice, minimum size=0.7cm, hair=purple, shirt=magenta,skin=pink, label=below:Alice]    at (-1.4,0.0) (10pt) {};
\node[bob,undershirt=black, mirrored, shirt=blue,minimum size=.7cm,label=below:Bob]   at (4.3,0) {};
\node[thick,black,label=below:Quantum state ($\varrho$)]   at (2,-0.2) {};
\node[thick,black,label=above: $|\psi\rangle$]   at (-0.6,0.01) {};
\draw[color=cyan, ultra thick] (-0.9,-0.3) rectangle (0.4,0.7);
\node[thick,black,label=above: BSM $\{|\phi_k\rangle\}$]   at (-0.2,0.6) {};
\node[ultra thick,black,label=above:\textbf{(c).}]   at (-2.4,-0.5) {};
\end{tikzpicture}
\end{minipage}%
\begin{minipage}{.5\textwidth}
  \centering
  \begin{tikzpicture}
\draw[snake=snake,red,ultra thick] (-0.8,-2) -- (0,-2);
\shade[ball color=blue] (4,-0.8) circle (.2cm);
\shade[ball color=orange!67] (4,-2.6) circle (.2cm);
\shade[ball color=green] (0,-2) circle (.2cm);
\shade[ball color=green] (-0.9,-2) circle (.2cm);
\draw[color=black, thick] (-1.3,-1.25) rectangle (0.55,0.3);
\draw[color=red,ultra thick] (3.3,-0.45) rectangle (4.7,-01.3);
\draw[color=cyan,ultra thick] (-1.2,-2.7) rectangle (0.3,-1.7);
\node[name=alice, shape=alice, minimum size=0.7cm, hair=purple, shirt=magenta,skin=pink, label=below:Alice]    at (-0.8,1.5) (10pt) {};
\node[bob,undershirt=black, mirrored, shirt=blue,minimum size=.7cm,label=below:Bob]   at (4.,1.5) {};
\node[thick,black,label=below:Noisy classical channel]   at (1.8,1.3) {};
\node[ultra thick,black,label=above:\textbf{(d).}]   at (-1.8,1) {};
\node[ultra thick,black,label=below:$|\phi_k\rangle$]   at (-0.4,-2) {};
\node[ultra thick,black,label=below:\phone]   at (-0.2,1.5) {};
\node[ultra thick,black,label=below:\phone]   at (3.6,1.5) {};
\draw[snake=zigzag,green,ultra thick] (-0.2,1.2) -- (3.6,1.2);

\draw[-{Implies},double,ultra thick,cyan] (4,0.3)-- (4,-0);
\draw[-{Implies},double,ultra thick,cyan] (-0.4,-1.6)-- (-0.4,-1.3);
\draw[-{Implies},double,ultra thick,cyan] (-0.2,0.4)-- (-0.2,0.7);
\draw[-{Implies},double,ultra thick,cyan] (4,-2)-- (4,-2.3);
\node[thick,black,label=below:{$k\in\{0,1,2,3\}$}]   at (-0.4,-2.6) {};
\node[thick,black,label=below:{$k=0\rightarrow00$}]   at (-0.4,0.2) {};
\node[thick,black,label=below:{$k=1\rightarrow11$}]   at (-0.4,-0.1) {};
\node[thick,black,label=below:{$k=2\rightarrow01$}]   at (-0.4,-0.4) {};
\node[thick,black,label=below:{$k=3\rightarrow10$}]   at (-0.4,-0.7) {};
\node[thick,black,label=below:{BSM$\rightarrow cbit$}]   at (-0.4,0.5) {};
\node[thick,black,label=above:{Unitary $(\varsigma_k)$}]   at (4,-0.6) {};
\node[thick,black,label=below:{$\varsigma_k\in\{I_2,\sigma_x,\sigma_y,\sigma_z\}$}]   at (4,-1.2) {};
\node[thick,black,label={$\varrho_k^{\prime}$}]   at (4.5,-3) {};
\end{tikzpicture}
\end{minipage}
\caption{These figures depict our teleportation protocol where Alice and Bob are spatially separated. (a). Alice and Bob share an arbitrary two-qubit state in canonical form between them. (b). Alice is given an unknown state $|\psi\rangle$, which she wants to teleport to Bob. (c). Alice measures her qubits in Bell basis, also known as the Bell states measurement (BSM). (d). After getting the measurement outcome, Alice sends it to Bob through noisy classical channel(s). Then, based on the measurement outcome received, Bob applies the corresponding Pauli's unitary operator to his qubit.}
\label{image}
\end{figure}
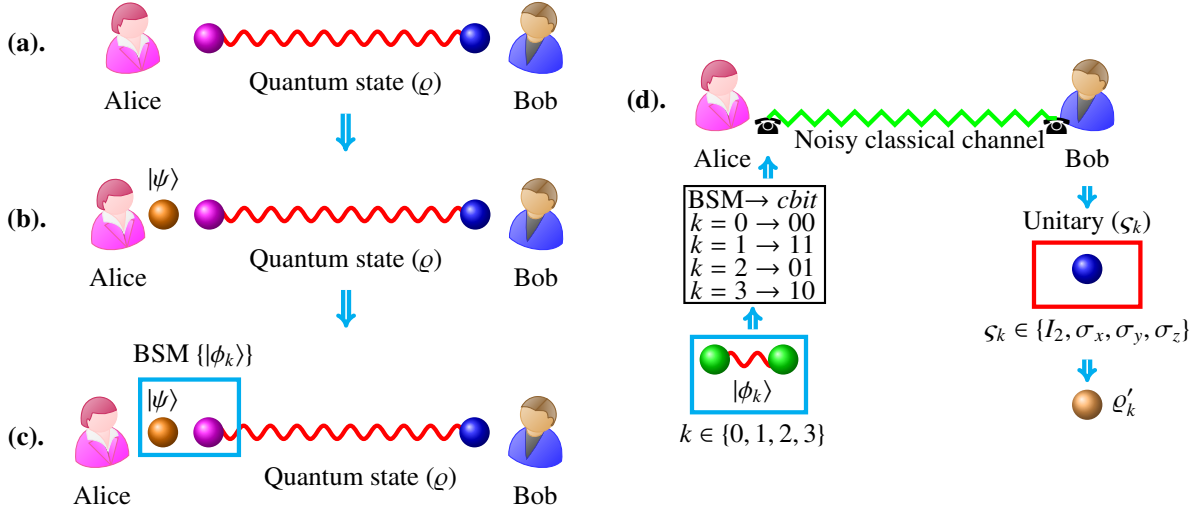

{\it Noise Model-I}: Let us first consider a single noisy classical channel, through which Alice tries to send two-cbit of classical information. This channel is characterized by the following conditional probabilities:
\begin{equation}\label{eq3}
\begin{array}{c}
P(ab|ab) = p_0,~ P(\bar{a}b|ab) = p_1,\\ 
P(a\bar{b}|ab) = p_2,~P(\bar{a}\bar{b}|ab) = p_3,
\end{array}
\end{equation}
where both $a$ and $b$ take the value either 0 or 1, $\bar{a}$ and $\bar{b}$ are the complements of $a$ and $b$ respectively. Alice's input is one of the bit strings, drawn from the set \{00, 01, 10, 11\}. The probabilities, $p_i\geq0$ for $\forall i = 0, \dots, 3$, $\sum_{i=0}^3 p_i = 1$. 

Notice that $p_0$ is the probability, responsible for appropriate unitary rotation and $p_1$, $p_2$, $p_3$ introduce error to Bob's choice of unitary after receiving the information regarding Alice's measurement outcome. 

It is possible to quantify the amount of classical information, conveyed by Alice to Bob. The quantification is based on the mutual information between Alice's input $X$ and Bob's output $Y$. It is given as the following:
\begin{equation}\label{eq4}
\mathbf{I}(X:Y) = 2+\sum_{i=0}^3p_i\log_2 (p_i).
\end{equation}

In the Table \ref{tab2}, we summarize the measurement outcomes, corresponding cbits of Alice, the cbits, received by Bob and the unitary rotations.

{\it Noise Model-II}: We now consider that Alice sends the measurement outcome using two independent noisy classical channels, i.e., one cbit through each classical channel. These classical channels are labeled by $C_l$, $l$ is either $\eta$ or $\eta^\prime$. The conditional probabilities are given as:
\begin{equation} \label{eq5}
P(a|a) = l,~P(\bar{a}|a) = 1-l,
\end{equation}
where $a$ is Alice’s input bit, takes the value 0 or 1, $\bar{a}$ is the complement of $a$. The values of $\eta$ and $\eta^{\prime}$ lie between $\frac{1}{2}$ to 1 i.e., $\frac{1}{2}\leq\eta,\eta^{\prime}\leq1$. Again, the amount of classical information which is communicated can be quantified using the concept of mutual information, given as the following:
\begin{equation}\label{eq6}
\mathbf{I}_l = 1 - H(l),
\end{equation}
where $H(l) = - l\log_2l - (1-l)\log_2(1-l)$, the binary Shannon entropy. 



We now define two important quantities the average fidelity corresponding to the state $\varrho$, i.e., $\mathbf{F}_{\varrho}$ and its deviation $\Delta_{\varrho}$. The average fidelity can be defined as the following:
\begin{equation}\label{eq7}
\mathbf{F}_\varrho = \langle\delta\rangle = \int f_\varrho~\mbox{d}\psi,
\end{equation}
where $f_\varrho$ is the fidelity between the unknown state which Alice wants to teleport (the input state) and the state which is prepared on Bob's side after his unitary operation (the output state). Obviously, $f_\varrho$ corresponds to the state $\varrho$ in the canonical form. The integral is taken over a uniform distribution $\mbox{d}\psi$ with respect to all input states, where $\int \mbox{d}\psi = 1$ (normalized Haar measure). 


We consider here fidelity deviation $\Delta_\varrho$ corresponding to $\mathbf{F}_\varrho$. It is defined as: 
\begin{equation}\label{eq8}
\Delta_\varrho = \sqrt{\langle\delta^2\rangle-\mathbf{F}_\varrho^2},
\end{equation}
where $\langle\delta^2\rangle$ = $\int f_\varrho^2~\mbox{d}\psi$. We want the above quantity to be minimum for a given resource state. In particular, if $\Delta_\varrho=0$ in the standard teleportation protocol, then the given resource state helps to teleport all input states equally well. But for the present protocol, it is not known how to minimize the fidelity deviation corresponding to an arbitrary resource state. However, we stick to the pair of quantities $\{\mathbf{F}_\varrho, \Delta_\varrho\}$ and examine the combined effects of noisy resource state along with noisy classical communication on these quantities while teleporting an unknown qubit. 


\begin{table}[h]
\begin{center}
\begin{tabular}{|c|c|c|c|c|}
\hline
Measurement & The Bell state & cbits, sent & cbits, received by Bob & Unitary operation\\
outcome & & by Alice & & \\
\hline
0 & $|\phi_0\rangle$ & 00 & 00 with $p_0$ & $\varsigma_0$ with $p_0$\\
   &                                   &       & 10 with $p_1$ & $\varsigma_3$ with $p_1$\\
   &                                   &       & 01 with $p_2$ & $\varsigma_2$ with $p_2$\\
   &                                   &       & 11 with $p_3$ & $\varsigma_1$ with $p_3$\\
\hline\hline
1 & $|\phi_1\rangle$ & 11 & 11 with $p_0$ & $\varsigma_1$ with $p_0$\\
   &                                   &       & 01 with $p_1$ & $\varsigma_2$ with $p_1$\\
   &                                   &       & 10 with $p_2$ & $\varsigma_3$ with $p_2$\\
   &                                   &       & 00 with $p_3$ & $\varsigma_0$ with $p_3$\\
\hline\hline
2 & $|\phi_2\rangle$ & 01 & 01 with $p_0$ & $\varsigma_2$ with $p_0$\\
   &                                   &       & 11 with $p_1$ & $\varsigma_1$ with $p_1$\\
   &                                   &       & 00 with $p_2$ & $\varsigma_0$ with $p_2$\\
   &                                   &       & 10 with $p_3$ & $\varsigma_3$ with $p_3$\\
\hline\hline
3 & $|\phi_3\rangle$ & 10 & 10 with $p_0$ & $\varsigma_3$ with $p_0$\\
   &                                   &       & 00 with $p_1$ & $\varsigma_0$ with $p_1$\\
   &                                   &       & 11 with $p_2$ & $\varsigma_1$ with $p_2$\\
   &                                   &       & 01 with $p_3$ & $\varsigma_2$ with $p_3$\\
\hline
\end{tabular}
\caption{In the above, we provide Alice's measurement outcomes, corresponding Bell states and Alice's cbits. Then, we provide the cbits which is received by Bob and the unitary operations, associated with certain probabilities.}\label{tab2}
\end{center}
\end{table}

We are now ready to present the main findings of the paper. 

\section{Results}\label{sec3}
{\bf Fidelity:} Let us consider the state which is to be teleported by Alice to bob, is $\ket{\psi} = \frac{1}{2}(\mathbb{I}_2 + \mbox{\bf a}\boldsymbol{\cdot\sigma})$ (Bloch sphere representation), {\bf a} is a unit vector in $\mathbb{R}^3$. After Alice's measurement in the Bell basis $\{\ket{\phi_k}\}_k$, $k=0,1,2,3$, [$|\phi_0\rangle = (1/\sqrt{2})(\ket{00}+\ket{11})$, $|\phi_1\rangle = (1/\sqrt{2})(\ket{00}-\ket{11})$, $|\phi_2\rangle = (1/\sqrt{2})(\ket{01}+\ket{10})$, and $|\phi_3\rangle = (1/\sqrt{2})(\ket{01}-\ket{10})$] the state which is prepared on Bob's side (before the application of the unitary operator), is given by- 
\begin{equation}\label{eq9}
\varrho_k=\frac{1}{s_k}\mbox{Tr}_{A}\left[(|\phi_k\rangle\langle\phi_k|\otimes\mathbb{I}_2)(|\psi\rangle\langle\psi|\otimes\varrho)(|\phi_k\rangle\langle\phi_k|\otimes\mathbb{I}_2)\right],
\end{equation}
where the resource state is $\varrho$, the trace is taken over Alice's two-qubit system. `$s_k$' is the probability of getting the outcome $k$, it can be defined as: $s_k = \mbox{Tr}\left[(|\phi_k\rangle\langle\phi_k|\otimes\mathbb{I}_2)(|\psi\rangle\langle\psi|\otimes\varrho)(|\phi_k\rangle\langle\phi_k|\otimes\mathbb{I}_2)\right]$. Using the Hilbert-Schmidt representations, $\varrho_k$ can be written as the following:
\begin{equation}\label{eq10}
\varrho_k=\frac{1}{8s_k}\left[(1+\mbox{\bf a}^T\mathbb{T}_k\mbox{\bf r})\mathbb{I}_2+(\mbox{\bf s}+\mathbb{T}^{\dagger}\mathbb{T}_k\mbox{\bf a})\boldsymbol{\cdot\sigma}\right],
\end{equation}
where the matrices ($\mathbb{T}_k$) correspond to the Bell states $|\phi_k\rangle$ $\forall k = 0,\dots,3$, they are given as: $\mathbb{T}_0 = \mbox{diag}(1,-1,1)$ for $|\phi_0\rangle$, $\mathbb{T}_1 = \mbox{diag}(-1,1,1)$ for $|\phi_1\rangle$, $\mathbb{T}_2 = \mbox{diag}(1,1,-1)$ for $|\phi_2\rangle$, $\mathbb{T}_3 = \mbox{diag}(-1,-1,-1)$ for $|\phi_3\rangle$, the matrix $\mathbb{T}$ and the vectors {\bf r, s} correspond to $\varrho$, $\mbox{\bf a}^T$ is produced by taking the transpose of the unit vector {\bf a}. Remember that originally, we have defined correlation matrix for any state $\rho$ but in our protocol we start with a state in the canonical form, and thus, we use the state $\varrho$. So, henceforth $\mathbb{T}$ belongs to $\varrho$, which is a $3\times3$ diagonal matrix. Now, after generation of $\varrho_k$, Bob applies the Pauli's unitary operator on it based on the classical message that he receives from Alice. The details of the application of unitary operators are given in Table \ref{tab2}. We follow the Noise Model-I first. The states which are produced after the application of the Pauli's unitary operators, are given by-
\begin{equation}\label{eq11}
\begin{array}{c}
\varrho^\prime_0 = p_0\varsigma_0\varrho_0\varsigma_0 + p_1\varsigma_3\varrho_0\varsigma_3 + p_2\varsigma_2\varrho_0\varsigma_2 + p_3\varsigma_1\varrho_0\varsigma_1,\\

\varrho^\prime_1 = p_0\varsigma_1\varrho_1\varsigma_1 + p_1\varsigma_2\varrho_1\varsigma_2 + p_2\varsigma_3\varrho_1\varsigma_3 + p_3\varsigma_0\varrho_1\varsigma_0,\\

\varrho^\prime_2 = p_0\varsigma_2\varrho_2\varsigma_2 + p_1\varsigma_1\varrho_2\varsigma_1 + p_2\varsigma_0\varrho_2\varsigma_0 + p_3\varsigma_3\varrho_2\varsigma_3,\\

\varrho^\prime_3 = p_0\varsigma_3\varrho_3\varsigma_3 + p_1\varsigma_0\varrho_3\varsigma_0 + p_2\varsigma_1\varrho_3\varsigma_1 + p_3\varsigma_2\varrho_3\varsigma_2,
\end{array}
\end{equation}
where each $\varrho_k^\prime$ corresponds to the outcome $k$. The unitary $\varsigma_k$ corresponds the outcome $k$. When the outcome $k$ is conveyed to Bob via noisy classical channel, Bob gets the original message with probability $p_0$ only. Thus, with other probabilities, Bob applies wrong unitary operators. In this way, the average state which is produced on bob's side, is given by-
\begin{equation}\label{eq12}
\begin{array}{l}
\varrho_{avg} = \sum_{k=0}^3s_k\varrho_k^{\prime} = \\
p_0\left(s_0\varsigma_0\varrho_0\varsigma_0 + s_1\varsigma_1\varrho_1\varsigma_1 + s_2\varsigma_2\varrho_2\varsigma_2 + s_3\varsigma_3\varrho_3\varsigma_3\right) +\\

p_1\left(s_0\varsigma_3\varrho_0\varsigma_3 + s_1\varsigma_2\varrho_1\varsigma_2 + s_2\varsigma_1\varrho_2\varsigma_1 + s_3\varsigma_0\varrho_3\varsigma_0\right) +\\

p_2\left(s_0\varsigma_2\varrho_0\varsigma_2 + s_1\varsigma_3\varrho_1\varsigma_3 + s_2\varsigma_0\varrho_2\varsigma_0 + s_3\varsigma_1\varrho_3\varsigma_1\right) +\\

p_3\left(s_0\varsigma_1\varrho_0\varsigma_1 + s_1\varsigma_0\varrho_1\varsigma_0 + s_2\varsigma_3\varrho_2\varsigma_3 + s_3\varsigma_2\varrho_3\varsigma_2\right).\\
\end{array}
\end{equation} 
From the above equation, we now consider the following term: $s_0\varsigma_0\varrho_0\varsigma_0 + s_1\varsigma_1\varrho_1\varsigma_1 + s_2\varsigma_2\varrho_2\varsigma_2 + s_3\varsigma_3\varrho_3\varsigma_3$. This term can be written as:
\begin{equation}\label{eq13}
\begin{array}{l}
\frac{1}{8}\left[(1+\mbox{\bf a}^T\mathbb{T}_0\mbox{\bf r})\mathbb{I}_2 + \mathcal{O}^{\dagger}_0(\mbox{\bf s}+ \mathbb{T}^{\dagger}\mathbb{T}_0\mbox{\bf a})\boldsymbol{\cdot\sigma} + (1+\mbox{\bf a}^T\mathbb{T}_1\mbox{\bf r})\mathbb{I}_2 + \mathcal{O}^{\dagger}_1(\mbox{\bf s} + \mathbb{T}^{\dagger}\mathbb{T}_1\mbox{\bf a})\boldsymbol{\cdot\sigma}\right]+\\[2 ex]

\frac{1}{8}\left[(1+\mbox{\bf a}^T\mathbb{T}_2\mbox{\bf r})\mathbb{I}_2+\mathcal{O}^{\dagger}_2(\mbox{\bf s}+\mathbb{T}^{\dagger}\mathbb{T}_2\mbox{\bf a})\boldsymbol{\cdot\sigma}+(1+\mbox{\bf a}^T\mathbb{T}_3\mbox{\bf r})\mathbb{I}_2 + \mathcal{O}^{\dagger}_3(\mbox{\bf s} + \mathbb{T}^{\dagger}\mathbb{T}_3\mbox{\bf a})\boldsymbol{\cdot\sigma}\right].
\end{array}
\end{equation}
In the above we use the following fact: $\varsigma_k(\mbox{\bf n}\boldsymbol{\cdot\sigma})\varsigma_k = (\mathcal{O}^{\dagger}_k\mbox{\bf n})\boldsymbol{\cdot\sigma}$, $\forall k = 0,\dots, 3$. Each $\mathcal{O}_k$ is a rotation in $\mathbb{R}^3$. Here $\mathcal{O}^{\dagger}_k = -\mathbb{T}_k$, $\forall k = 0,\dots,3$ (for a greater details one can go through the Refs.~\cite{Badziag00, Ghosal20-1}). Notice that $\mathbb{T}^\dagger$ and $\mathbb{T}_k$ commutes with each other for all values of $k$ and thus, we use the relation $\mathbb{T}^{\dagger}\mathbb{T}_k = \mathbb{T}_k\mathbb{T}^{\dagger}$ in the above expression, i.e., in Eq.~(\ref{eq13}). Moreover, we also use the relations $\sum_{k=0}^3\mathbb{T}_k = 0$ and $\mathbb{T}_k^2 = \mathbb{I}_3$, $\forall k = 0,\dots,3$ in Eq.~(\ref{eq13}). We finally get $s_0\varsigma_0\varrho_0\varsigma_0 + s_1\varsigma_1\varrho_1\varsigma_1 + s_2\varsigma_2\varrho_2\varsigma_2 + s_3\varsigma_3\varrho_3\varsigma_3$ = $\frac{1}{2}(\mathbb{I}_2-\mathbb{T}^{\dagger}\mbox{\bf a}\boldsymbol{\cdot\sigma})$ = $\frac{1}{2}(\mathbb{I}_2 + \mathbb{T}^{\dagger}\mathbb{T}_3\mbox{\bf a}\boldsymbol{\cdot\sigma})$, remembering the fact that $\mathbb{T}_3 = -\mathbb{I}_3$. Following the similar steps and using the relations $\mathbb{T}_0\mathbb{T}_3$ = $\mathbb{T}_3\mathbb{T}_0$ = $-\mathbb{T}_0$ = $\mathbb{T}_2\mathbb{T}_1$ = $\mathbb{T}_1\mathbb{T}_2$, it is possible to show that $s_0\varsigma_3\varrho_0\varsigma_3 + s_1\varsigma_2\varrho_1\varsigma_2 + s_2\varsigma_1\varrho_2\varsigma_1 + s_3\varsigma_0\varrho_3\varsigma_0$ = $\frac{1}{2}\left(\mathbb{I}_2 + \mathbb{T}^{\dagger}\mathbb{T}_0\mbox{\bf a}\boldsymbol{\cdot\sigma}\right)$. Again, it is also possible to show that $s_0\varsigma_2\varrho_0\varsigma_2 + s_1\varsigma_3\varrho_1\varsigma_3 + s_2\varsigma_0\varrho_2\varsigma_0 + s_3\varsigma_1\varrho_3\varsigma_1$ = $\frac{1}{2}\left(\mathbb{I}_2+\mathbb{T}^{\dagger}\mathbb{T}_1\mbox{\bf a}\boldsymbol{\cdot\sigma}\right)$, using the relations $\mathbb{T}_0\mathbb{T}_2$ = $\mathbb{T}_2\mathbb{T}_0 $ = $-\mathbb{T}_1$ = $\mathbb{T}_3\mathbb{T}_1$ = $\mathbb{T}_1\mathbb{T}_3$ and $s_0\varsigma_1\varrho_0\varsigma_1 + s_1\varsigma_0\varrho_1\varsigma_0 + s_2\varsigma_3\varrho_2\varsigma_3 + s_3\varsigma_2\varrho_3\varsigma_2$ = $\frac{1}{2}\left(\mathbb{I}_2+\mathbb{T}^{\dagger}\mathbb{T}_2\mbox{\bf a}\boldsymbol{\cdot\sigma}\right)$, using the relations $\mathbb{T}_0\mathbb{T}_1$ = $\mathbb{T}_1\mathbb{T}_0$ = $-\mathbb{T}_2$ = $\mathbb{T}_3\mathbb{T}_2$ = $\mathbb{T}_2\mathbb{T}_3$. Clearly, $\varrho_{avg}$ can be rewritten in the following way:
\begin{equation}\label{eq14}
\begin{array}{l}
\varrho_{avg} = \frac{1}{2}\left(\mathbb{I}_2 + \mathbb{T}^{\dagger}(p_0\mathbb{T}_3 + p_1\mathbb{T}_0 + p_2\mathbb{T}_1 + p_3\mathbb{T}_2)\mbox{\bf a}\boldsymbol{\cdot\sigma}\right)= \frac{1}{2}\left(\mathbb{I}+\mathcal{X}\mbox{\bf a}\boldsymbol{\cdot\sigma}\right),
\end{array}
\end{equation}
where $\sum_{i = 0}^3p_i = 1$ and the matrix $\mathcal{X} = \mathbb{T}^{\dagger}(p_0\mathbb{T}_3 + p_1\mathbb{T}_0 + p_2\mathbb{T}_1 + p_3\mathbb{T}_2)$ is a $3\times 3$ real diagonal matrix because the correlation matrices are $3\times3$ real diagonal matrices. Using $\varrho_{avg}$, we calculate $f_\varrho$ which is given by-
\begin{equation}\label{eq15}
\begin{array}{l}
f_\varrho = \mbox{Tr}\left(\varrho_{avg}|\psi\rangle\langle\psi|\right)= \frac{1}{2}\left(1+\mbox{\bf a}^T\mathcal{X}\mbox{\bf a}\right),
\end{array}
\end{equation}
where $|\psi\rangle\langle\psi| = \frac{1}{2}\left(\mathbb{I}_2 + \mbox{\bf a}\boldsymbol{\cdot\sigma}\right)$ is the unknown input state which Alice wants to teleport. We recall that we want to calculate the average fidelity $\mathbf{F}_\varrho$, corresponding to the resource state $\varrho$ which is in the canonical form. For this purpose, we apply Schur's lemma on $\mathbb{R}^d$ (for a detailed description of this lemma one can go through the Refs.~\cite{Horodecki96, Bang18, Ghosal20-1} and the references therein). This lemma leads us to the expression of $\mathbf{F}_\varrho$, given by-
\begin{equation}\label{eq16}
\mathbf{F}_\varrho = \int f_{\varrho} \mbox{d{\bf a}} = \int\frac{\left(1+\mbox{\bf a}^{T}\mathcal{X}\mbox{\bf a}\right)}{2}\mbox{d{\bf a}} = \frac{1}{2}\left(1+\frac{\mbox{Tr}(\mathcal{X})}{3}\right),
\end{equation}
where we use $\int\mbox{\bf a}^T\mathcal{X}\mbox{\bf a}~\mbox{d{\bf a}} = \frac{1}{3}\mbox{Tr}(\mathcal{X})$ by Schur's lemma. The above expression is the desired expression for teleportation fidelity for our protocol, starting from an arbitrary two-qubit state $\rho$ when we use the Noise Model-I for noisy classical communication. From Eq.~(\ref{eq15}) and Eq.~(\ref{eq16}), it is clear that if classical communication is noiseless that means Bob applies the right unitary operation ($p_0=1$) after getting the message from Alice then $f_{\varrho}$  and $F_{\varrho}$ can be written as $f_{\varrho}=\frac{1}{2}(1-a^TTa)$ , $F_{\varrho}=\frac{1}{2}\big(1-\frac{\mbox{Tr}T}{3}\big)$ respectively which are consistent with the results of Refs.~\cite{Badziag00, Ghosal20-1}. 
So, the present expressions provides us the idea regarding the combined effect of noisy resource state along with noisy classical communication on the average value of the teleportation fidelity while teleporting an unknown qubit.

Using $\mathbb{T}$ = diag($\lambda_i|\mathbf{t}_{ii}|$, $\lambda_j|\mathbf{t}_{jj}|$, $\lambda_k|\mathbf{t}_{kk}|$) and the forms of $\mathbb{T}_k$, we calculate $\mbox{Tr}(\mathcal{X})$ which is given as: 
\begin{equation}\label{eq17}
\begin{array}{l}
\mbox{Tr}(\mathcal{X})=\left(p_1+p_3-p_0-p_2\right)\lambda_i|\mathbf{t}_{ii}|+\left(p_2+p_3-p_0-p_1\right)\lambda_j|\mathbf{t}_{jj}|\\
~~~~~~~~+\left(p_1+p_2-p_0-p_3\right)\lambda_k|\mathbf{t}_{kk}|,
\end{array}
\end{equation}
where $i\neq j\neq k\in\{1,2,3\}$. Now, rearranging the above and using the relation $\sum_{i=0}^3p_i=1$, we get- 
\begin{equation} \label{eq18}
\begin{array}{l}
\mbox{Tr}(\mathcal{X})=2p_1\left(\lambda_i|\mathbf{t}_{ii}|+\lambda_k|\mathbf{t}_{kk}|\right)+2p_2\left(\lambda_j|\mathbf{t}_{jj}|+\lambda_k|\mathbf{t}_{kk}|\right)\\
~~~~~~~~+2p_3\left(\lambda_i|\mathbf{t}_{ii}|+\lambda_j|\mathbf{t}_{jj}|\right)-\sum_{i=1}^3\lambda_i|\mathbf{t}_{ii}|.
\end{array}
\end{equation}
We recall the discussion, given just after Eq.~(\ref{eq2}). From that discussion, if det$\mathbb{T}\leq0$ then we can take $\lambda_i = \lambda_j = \lambda_k = -1$ for all $|\mathbf{t}_{ii}|\neq0$, $i=1,2,3$ and $\mathbb{T}$ = diag($-|\mathbf{t}_{11}|$, $-|\mathbf{t}_{22}|$, $-|\mathbf{t}_{33}|$). One can also see Ref.~\cite{Ghosal20-1} in this regard. Here the idea is to provide an easily calculable formula for the average value of the teleportation fidelity. So, using these values of $\lambda_i$, $\lambda_j$, and $\lambda_k$, we get an expression of $\mathbf{F}_\varrho$, which is given as
\begin{equation}\label{eq19}
\begin{array}{l}
\mathbf{F}_{\varrho} = \frac{1}{2}\left(1+\frac{1}{3}\sum_{i=1}^3|\mathbf{t}_{ii}|\right)\\-\frac{1}{3}\Bigg(p_1\left(|\mathbf{t}_{11}|+|\mathbf{t}_{33}|\right)+p_2\left(|\mathbf{t} _{22}|+|\mathbf{t}_{33}|\right)+p_3\left(|\mathbf{t}_{11}|+|\mathbf{t}_{22}|\right)\Bigg),
\end{array}
\end{equation} 
where the term $\frac{1}{2}\left(1+\frac{1}{3}\sum_{i=1}^3|\mathbf{t}_{ii}|\right)$ is the teleportation fidelity for some two-qubit resource states when the classical communication is noiseless. We next consider det$\mathbb{T}>0$, i.e., $\lambda_i = \lambda_j = -1$, $\lambda_k = 1$ for any choice of $i\neq j\neq k\in\{1,2,3\}$ satisfying the condition $|\mathbf{t}_{ii}|\geq|\mathbf{t}_{jj}|\geq|\mathbf{t}_{kk}|$. Hence, $\mathbf{F}_\varrho$ becomes:
\begin{equation}\label{eq20}
\begin{array}{l}
\mathbf{F}_{\varrho} = \frac{1}{2}\left(1+\frac{1}{3}(|\mathbf{t}_{ii}|+|\mathbf{t}_{jj}|-|\mathbf{t}_{kk}|)\right)\\-\frac{1}{3}\Bigg(p_1\left(|\mathbf{t}_{ii}|-|\mathbf{t}_{kk}|\right)+p_2\left(|\mathbf{t} _{jj}|-|\mathbf{t}_{kk}|\right)+p_3\left(|\mathbf{t}_{ii}|+|\mathbf{t}_{jj}|\right)\Bigg).
\end{array}
\end{equation} 
Remember that the indices $i,j,k$ are determined, maintaining the ordering $|\mathbf{t}_{ii}|\geq|\mathbf{t}_{jj}|\geq|\mathbf{t}_{kk}|$. Thus, the term $\frac{1}{2}\left(1+\frac{1}{3}(|\mathbf{t}_{ii}|+|\mathbf{t}_{jj}|-|\mathbf{t}_{kk}|)\right)$ is maximized. This maximization is required to find the value of $\mathbf{F}_{\varrho}$. 

Here we make some important points: (i) from Eq.~(\ref{eq19}) and from Eq.~(\ref{eq20}), it is clear that $\mathbf{F}_{\varrho}$ becomes the optimal teleportation fidelity in the standard teleportation protocol where the resource state is an arbitrary two-qubit state and when $p_1 = p_2 = p_3 = 0$, implying the fact that the classical communication is noiseless and that is consistent with Refs.~\cite{Badziag00, Ghosal20-1}, (ii) the term, $\frac{1}{3}(p_1\left(|\mathbf{t}_{11}|+|\mathbf{t}_{33}|\right)+p_2\left(|\mathbf{t} _{22}|+|\mathbf{t}_{33}|\right)+p_3\left(|\mathbf{t}_{11}|+|\mathbf{t}_{22}|\right))$ is due to the noise, associated with the classical communication. We say this term as $f_{noise}/3$ and using this term we find the condition for non-classical fidelity for our protocol when det$\mathbb{T}<0$. We take here the definition of non-classical fidelity as the fidelity which is greater than the maximum fidelity ($\frac{2}{3}$), achievable through the classical teleportation \cite{Popescu94, Massar95}. However, we consider only those states for which det$\mathbb{T}$ is less than zero (this includes all mixed entangled states as well) because it is known that in the standard teleportation protocol a resource state can lead to non-classical fidelity only when det$\mathbb{T}<0$, i.e., $\sum_{i=1}^3|\mathbf{t}_{ii}|>1$ (see Refs.~\cite{Popescu94, Massar95, Horodecki96, Horodecki96-1}). In the present case also, we use the standard teleportation protocol but here the resource state can be an arbitrary two-qubit entangled state and there is also noise in the classical communication. This is why the condition for non-classical fidelity gets modified here. The modified condition is given by-
\begin{equation}\label{eq21}
\begin{array}{l}
\frac{1}{2}(1+\frac{1}{3}\sum_{i=1}^3|\mathbf{t}_{ii}|)-\frac{1}{3}f_{noise}>\frac{2}{3}~~
\Rightarrow ~~\sum_{i=1}^3|\mathbf{t}_{ii}|>(1+2f_{noise}).
\end{array}
\end{equation}
Within our protocol, using Noise Model-I, if a resource state satisfies the above condition, than that state leads to non-classical fidelity. Notice that the quantity $f_{noise}$ is a nonzero positive number and therefore, not all entangled states satisfy the above condition.

We also find out expressions for teleportation fidelity using Noise Model-II in our protocol. For this purpose, we use the following transformation rules: $p_0\rightarrow \eta\eta^\prime, p_1\rightarrow(1-\eta)\eta^\prime, p_2\rightarrow\eta(1-\eta^\prime), p_3\rightarrow(1-\eta)(1-\eta^\prime)$. Using these transformation rules, along with Eq.~(\ref{eq16}), we get the following expression for teleportation fidelity:
\begin{equation}\label{eq22}
\mathbf{F}_{\varrho} = \frac{1}{2}\left(1+\frac{\mbox{Tr}(\mathcal{X^\prime})}{3}\right),
\end{equation}
where $\mathcal{X^\prime}$ = $\mathbb{T}^{\dagger}[\eta\eta^\prime\mathbb{T}_3+(1-\eta)\eta^\prime\mathbb{T}_0+\eta(1-\eta^\prime)\mathbb{T}_1+(1-\eta)(1-\eta^\prime)\mathbb{T}_2]$ and $\mathbb{T}$, $\mathbb{T}_k$ have their usual meaning $\forall k=0,\dots,3$. Using $\mathbb{T}$ = diag($\lambda_i|\mathbf{t}_{ii}|$, $\lambda_j|\mathbf{t}_{jj}|$, $\lambda_k|\mathbf{t}_{kk}|$) and the forms of $\mathbb{T}_k$, we calculate $\mbox{Tr}(\mathcal{X}^\prime)$ which is given as: 
\begin{equation}\label{eq23}
\begin{array}{l}
\mbox{Tr}(\mathcal{X}^\prime)=\left(1-2\eta\right)\lambda_i|\mathbf{t}_{ii}|+\left(1-2\eta^\prime\right)\lambda_j|\mathbf{t}_{jj}|-\left(1-2\eta\right)\left(1-2\eta^\prime\right)\lambda_k|\mathbf{t}_{kk}|,
\end{array}
\end{equation}
where $i\neq j\neq k\in\{1,2,3\}$. For det$\mathbb{T}\leq0$, we take $\lambda_i=\lambda_j=\lambda_k=-1$ for all $|\mathbf{t}_{ii}|\neq0$, $i=1,2,3$ and $\mathbb{T}$ = diag($-|\mathbf{t}_{11}|$, $-|\mathbf{t}_{22}|$, $-|\mathbf{t}_{33}|$). Hence, $\mathbf{F}_{\varrho}$ becomes
\begin{equation}\label{eq24}
\begin{array}{r}
\mathbf{F}_{\varrho}=\frac{1}{2}(1+\frac{1}{3}\sum_{i=1}^3|\mathbf{t}_{ii}|)-\frac{1}{3}[|\mathbf{t}_{11}|(1-\eta)+|\mathbf{t}_{22}|(1-\eta^\prime)+|\mathbf{t}_{33}|(\eta+\eta^\prime-2\eta\eta^\prime)].
\end{array}   
\end{equation}
On the other hand, if det$\mathbb{T}>0$, then $\lambda_i=\lambda_j=-1$, $\lambda_k=1$ for any choice of $i\neq j\neq k\in\{1,2,3\}$ satisfying the condition $|t_{ii}|\geq|t_{jj}|\geq|t_{kk}|$. So, $\mathbf{F}_{\varrho}$ becomes
\begin{equation}\label{eq25}
\begin{array}{l}
\mathbf{F}_{\varrho}=\frac{1}{2}[1+\frac{1}{3}(|\mathbf{t}_{ii}|+|\mathbf{t}_{jj}|-|\mathbf{t}_{kk}|)]\\-\frac{1}{3}[|\mathbf{t}_{ii}|(1-\eta)+|\mathbf{t}_{jj}|(1-\eta^\prime)-|\mathbf{t}_{kk}|(\eta+\eta^\prime-2\eta\eta^\prime)].
\end{array}
\end{equation}
Remember that the indices $i,j,k$ are determined in such a way that the ordering $|\mathbf{t}_{ii}|\geq|\mathbf{t}_{jj}|\geq|\mathbf{t}_{kk}|$ is maintained. Thus, the term $\frac{1}{2}[1+\frac{1}{3}(|\mathbf{t}_{ii}|+|\mathbf{t}_{jj}|-|\mathbf{t}_{kk}|)]$ is maximized. This maximization is required to find the value of $\mathbf{F}_{\varrho}$. However, from Eq.~(\ref{eq24}), it is clear that $\mathbf{F}_{\varrho}$ is equal to the teleportation fidelity in the standard teleportation protocol with the resource state $\varrho$, when $\eta=\eta^\prime=1$.

We next find out the condition for non-classical teleportation fidelity. For this purpose, we consider only those states for which det$\mathbb{T}<0$, i.e., $\sum_{i=1}^3|\mathbf{t}_{ii}|>0$. We use these states as they are useful in the standard teleportation protocol (discussed earlier). We say the term $[|\mathbf{t}_{11}|(1-\eta)+|\mathbf{t}_{22}|(1-\eta^\prime)+|\mathbf{t}_{33}|(\eta+\eta^\prime-2\eta\eta^\prime)]$ as $f_{noise}^\prime$ and using this term we find out the condition, given as the following:
\begin{equation}\label{eq26}
\begin{array}{l}
\frac{1}{2}(1+\frac{1}{3}\sum_{i=1}^3|\mathbf{t}_{ii}|)-\frac{1}{3}f_{noise}^\prime>\frac{2}{3},~~
\Rightarrow ~~\sum_{i=1}^3|\mathbf{t}_{ii}|>(1+2f_{noise}^\prime).
\end{array}
\end{equation}
Remember that in Noise Model-II, $\eta\eta^\prime$ is greater than or equal to the other probabilities $(1-\eta)\eta^\prime$, $\eta(1-\eta^\prime)$, $(1-\eta)(1-\eta^\prime)$ as $\frac{1}{2}\leq\eta,\eta^\prime\leq1$. We now head to calculate fidelity deviation for different noise models.
\vskip 0.1 in

{\bf Fidelity deviation:} The expression for fidelity deviation, $\Delta_{\varrho}$, corresponding to the fidelity $\mathbf{F}_{\varrho}$, is given in Eq.~(\ref{eq8}). It has been pointed out in Ref.~\cite{Ghosal20-1} that $\Delta_{\varrho}$ lies between 0 and $\frac{1}{2}$, i.e., $0\leq\Delta_{\varrho}\leq\frac{1}{2}$. In fact, $\Delta_{\varrho}=0$ if and only if $\mathbf{F}_{\varrho}=f_{\varrho}$ for all input states $\ket{\psi}$. The main motivation to study fidelity deviation is that the pair $\{\mathbf{F}_{\varrho}, \Delta_{\varrho}\}$ gives more information regarding quantum teleportation than just teleportation fidelity \cite{Bang18}. In order to calculate $\Delta_{\varrho}$ (we use Noise Model-I first), we first calculate $\langle\delta^2\rangle$. From Eq.~(\ref{eq7}) and Eq.~(\ref{eq16}), we get $\langle\delta^2\rangle$ = $\int f^2_{\varrho}\mbox{d{\bf a}}$, where $f_{\varrho}=\frac{1}{2}\left(1+\mbox{\bf a}^T\mathcal{X}\mbox{\bf a}\right)$. So, $\langle\delta^2\rangle$ can be written as $\int \frac{1}{4}\left[1+2\mbox{\bf a}^T\mathcal{X}\mbox{\bf a}+(\mbox{\bf a}^T\mathcal{X}\mbox{\bf a})(\mbox{\bf a}^T\mathcal{X}\mbox{\bf a})\right]\mbox{d{\bf a}}$. Here, $\int (1+2\mbox{\bf a}^T\mathcal{X}\mbox{\bf a})\mbox{d{\bf a}}$ = $(1+\frac{2\mbox{Tr}(\mathcal{X})}{3})$, following Eq.~(\ref{eq16}). To get the value of $\int (\mbox{\bf a}^T\mathcal{X}\mbox{\bf a})(\mbox{\bf a}^T\mathcal{X}\mbox{\bf a})\mbox{d{\bf a}}$, we use the generalization of Schur's lemma on $\mathbb{R}^d\otimes\mathbb{R}^d$ \cite{Bang18} and we get $\int (\mbox{\bf a}^T\mathcal{X}\mbox{\bf a})(\mbox{\bf a}^T\mathcal{X}\mbox{\bf a})\mbox{d{\bf a}}$ = $\int (\mbox{\bf a}^T\otimes\mbox{\bf a}^T)(\mathcal{X}\otimes\mathcal{X})(\mbox{\bf a}\otimes\mbox{\bf a})\mbox{d{\bf a}}$ = $\frac{1}{15}[(\mbox{Tr}(\mathcal{X}))^2+\mbox{Tr}(\mathcal{X}\mathcal{X}^{\dagger})+\mbox{Tr}(\mathcal{X}^2)]$. Clearly, $\langle\delta^2\rangle$ can be written as the following:
\begin{equation}\label{eq27}
\footnotesize
\langle\delta^2\rangle = \frac{1}{4}\left[1+\frac{2\mbox{Tr}(\mathcal{X})}{3}+\frac{(\mbox{Tr}(\mathcal{X}))^2+\mbox{Tr}(\mathcal{X}\mathcal{X}^{\dagger})+\mbox{Tr}(\mathcal{X}^2)}{15}\right].
\end{equation}
On the other hand, $\mathbf{F}_{\varrho}^2$ is given as the following:
\begin{equation}\label{eq28}
\footnotesize
\mathbf{F}_{\varrho}^2 = \frac{1}{4}\left(1+\frac{\mbox{Tr}(\mathcal{X})}{3}\right)^2 = \frac{1}{4}\left(1+\frac{2\mbox{Tr}(\mathcal{X})}{3}+\frac{(\mbox{Tr}(\mathcal{X}))^2}{9}\right).
\end{equation}
From the above two equations, we can easily compute $\Delta_{\varrho}$, given by-
\begin{equation}\label{eq29}
\Delta_{\varrho} = \sqrt{\langle\delta^2\rangle-\mathbf{F}_{\varrho}^2} = \sqrt{\frac{1}{30}\left(\mbox{Tr}(\mathcal{X}^2)-\frac{(\mbox{Tr}(\mathcal{X}))^2}{3}\right)}.
\end{equation}
The above is the desired formula for fidelity deviation when we use Noise Model-I. A similar expression can be obtained if we use Noise Model-II. It is given by- 
\begin{equation}\label{eq30}
\small
\Delta_{\varrho} = \sqrt{\langle\delta^2\rangle-\mathbf{F}_{\varrho}^2} = \sqrt{\frac{1}{30}\left(\mbox{Tr}(\mathcal{X}^{\prime2})-\frac{(\mbox{Tr}(\mathcal{X}^\prime))^2}{3}\right)}.
\end{equation}
Using the explicit forms of $\mathcal{X}$ and $\mathcal{X}^\prime$, we can rewrite the expressions of $\Delta_{\varrho}$ for det$\mathbb{T}\leq0$ and for det$\mathbb{T}>0$. We first consider $\mathcal{X}$ = $\mathbb{T}^{\dagger}(p_0\mathbb{T}_3 + p_1\mathbb{T}_0 + p_2\mathbb{T}_1 + p_3\mathbb{T}_2)$. Considering $\mathcal{X}$, if det$\mathbb{T}\leq0$, then we define 
\begin{equation}\label{}
\Delta_{\varrho} = \left[1/30\left(\mathcal{A}_0-\mathcal{A}_1\right)\right]^{1/2},
\end{equation} 
and if det$\mathbb{T}>0$, then we define 
\begin{equation}\label{}
\Delta_{\varrho} = \left[1/30\left(\mathcal{A}_2 -\mathcal{A}_3\right)\right]^{1/2}. 
\end{equation}
$\mathcal{A}_i$, for all $i=0,\dots,3$ can be given in the following way: 
\begin{eqnarray}\label{}
\mathcal{A}_0 &=& (p_0+p_2-p_1-p_3)^2|\mathbf{t}_{11}|^2+(p_0+p_1-p_2-p_3)^2|\mathbf{t}_{22}|^2 +(p_0+p_3-p_1-p_2)^2|\mathbf{t}_{33}|^2, \nonumber\\ 
\mathcal{A}_1 &=& \frac{1}{3} \left[(p_0+p_2-p_1-p_3)|\mathbf{t}_{11}|+(p_0+p_1-p_2-p_3)|\mathbf{t}_{22}|  +(p_0+p_3-p_1-p_2)|\mathbf{t}_{33}|\right]^2,\\ 
\mathcal{A}_2 &=& (p_0+p_2-p_1-p_3)^2|\mathbf{t}_{ii}|^2+(p_0+p_1-p_2-p_3)^2|\mathbf{t}_{jj}|^2  +(p_0+p_3-p_1-p_2)^2|\mathbf{t}_{kk}|^2, \nonumber\\ 
\mathcal{A}_3 &=& \frac{1}{3} \left[(p_1+p_3-p_0-p_2)|\mathbf{t}_{ii}|+(p_2+p_3-p_0-p_1)|\mathbf{t}_{jj}| +(p_0+p_3-p_1-p_2)|\mathbf{t}_{kk}|\right]^2. \nonumber
\end{eqnarray}
In case of $\mathcal{A}_2$ and $\mathcal{A}_3$, the indices $i,j,k$ are determined in such a way that the ordering $|\mathbf{t}_{ii}|\geq|\mathbf{t}_{jj}|\geq|\mathbf{t}_{kk}|$ is maintained, for $i\neq j\neq k\in\{1,2,3\}$. Similarly, considering $\mathcal{X}^\prime$, we can rewrite the expressions for $\Delta_{\varrho}$ separately when det$\mathbb{T}\leq0$ and when det$\mathbb{T}>0$. For that purpose, we just have to use the transformation rules: $p_0\rightarrow \eta\eta^\prime, p_1\rightarrow(1-\eta)\eta^\prime, p_2\rightarrow\eta(1-\eta^\prime), p_3\rightarrow(1-\eta)(1-\eta^\prime)$ in the expressions of $\mathcal{A}_i$, $i=0,\dots,3$. We have mentioned earlier that we are mainly interested in the states for which det$\mathbb{T}<0$. These states can lead to zero fidelity deviation under Noise Model-I iff $\mathcal{A}_0$ = $\mathcal{A}_1$ which implies the following:
\begin{equation}\label{eq31}
\begin{array}{r}
|\mathbf{t}_{11}|[1-2(p_1+p_3)] = |\mathbf{t}_{22}|[1-2(p_2+p_3)] = |\mathbf{t}_{33}|[1-2(p_1+p_2)].
\end{array}
\end{equation}
Similarly, the states with det$\mathbb{T}<0$, can lead to zero fidelity deviation under Noise Model-II, iff the following condition is satisfied:
\begin{equation}\label{eq32}
\begin{array}{r}
|\mathbf{t}_{11}|(2\eta-1) = |\mathbf{t}_{22}|(2\eta^\prime-1)= |\mathbf{t}_{33}|(2\eta-1)(2\eta^\prime-1).
\end{array}
\end{equation}
From the analysis so far, we can summarize that (i) if a given two-qubit resource state obeys both Eq.~(\ref{eq21}) and Eq.~(\ref{eq31}), then such a state can provide both non-classical teleportation fidelity and zero fidelity deviation in our protocol under Noise Model-I, similarly, (ii) if a given two-qubit resource state obeys both Eq.~(\ref{eq26}) and Eq.~(\ref{eq32}), then such a state can provide both non-classical teleportation fidelity and zero fidelity deviation in our protocol under Noise Model-II. We note here that in a teleportation protocol these states are more desirable states than the states which can lead to non-classical fidelity but cannot lead to zero fidelity deviation \cite{Bang18, Ghosal20-1}. We like to mention here that in Eqs.~(\ref{eq31}) and (\ref{eq32}), we provide the conditions when fidelity deviation is zero. Here, these two equations depend upon the classical noise parameters and the resource state parameters. These conditions are more general and realistic conditions in a sense, if noisy classical channel parameters and resource state parameters both satisfy these equations then only we can have zero fidelity deviation. In general, one cannot control the classical noise parameters in realistic situations. But for a particular resource state one may get a classical channel where these equations will be satisfied then the deviation will be zero. Already we have seen that due to classical noise, teleportation fidelity gets reduced by some amount however, if someone is interested about only non-classical fidelity and zero deviation then he (or she) can think about implementing our protocol. 
\vskip 0.1 in

{\bf Several important aspects:} From Eq.~(\ref{eq31}), it is evident that even if $|\mathbf{t}_{11}| = |\mathbf{t}_{22}| = |\mathbf{t}_{33}|$, we may not get zero fidelity deviation (considering Noise Model-I). This is because the condition for zero fidelity deviation is also dependent on $p_1$, $p_2$, and $p_3$ along with $|\mathbf{t}_{11}|$, $|\mathbf{t}_{22}|$, $|\mathbf{t}_{33}|$. In the present teleportation protocol if there is no noise in the classical communication then a state can lead to zero fidelity deviation when $|\mathbf{t}_{11}| = |\mathbf{t}_{22}| = |\mathbf{t}_{33}|$ \cite{Ghosal20-1}. Such states can lead to zero fidelity deviation in the same protocol with noisy classical communication when $p_1=p_2=p_3$. So, there may exist a classical channel where this condition holds. On the other hand, if $p_1\neq p_2\neq p_3$ then the states with $|\mathbf{t}_{11}| = |\mathbf{t}_{22}| = |\mathbf{t}_{33}|$ must not lead to zero fidelity deviation. More importantly, the states which are not dispersion-free (i.e., cannot lead to zero fidelity deviation) in the present teleportation protocol with noiseless classical communication, can be dispersion-free (i.e., can lead to zero fidelity deviation) in the same protocol with noisy classical communication. Explicit example of which is given in a later portion. 

For Noise Model-II, if $|\mathbf{t}_{11}| = |\mathbf{t}_{22}| = |\mathbf{t}_{33}|$ for a given state, then from Eq.~(\ref{eq32}), we get $\eta = \eta^\prime = 1$ to make fidelity deviation zero (we exclude the choice $\eta=\eta^\prime=1/2$ as it implies that the amount of classical information communicated by Alice to Bob, is zero). But $\eta=\eta^\prime=1$ means classical channels are noiseless. Therefore, in the range $\frac{1}{2}<\eta,\eta^\prime<1$, we cannot make fidelity deviation zero for a state with $|\mathbf{t}_{11}| = |\mathbf{t}_{22}| = |\mathbf{t}_{33}|$. Clearly, Noise Model-I is comparatively better than the Noise Model-II when a state with $|\mathbf{t}_{11}| = |\mathbf{t}_{22}| = |\mathbf{t}_{33}|$, is given. We next consider some well-known two-qubit entangled states (det$\mathbb{T}<0$) and calculate the teleportation fidelity as well as fidelity deviation for those states.

\begin{figure}[h!]
\begin{center}
\includegraphics[width=.6\textwidth]{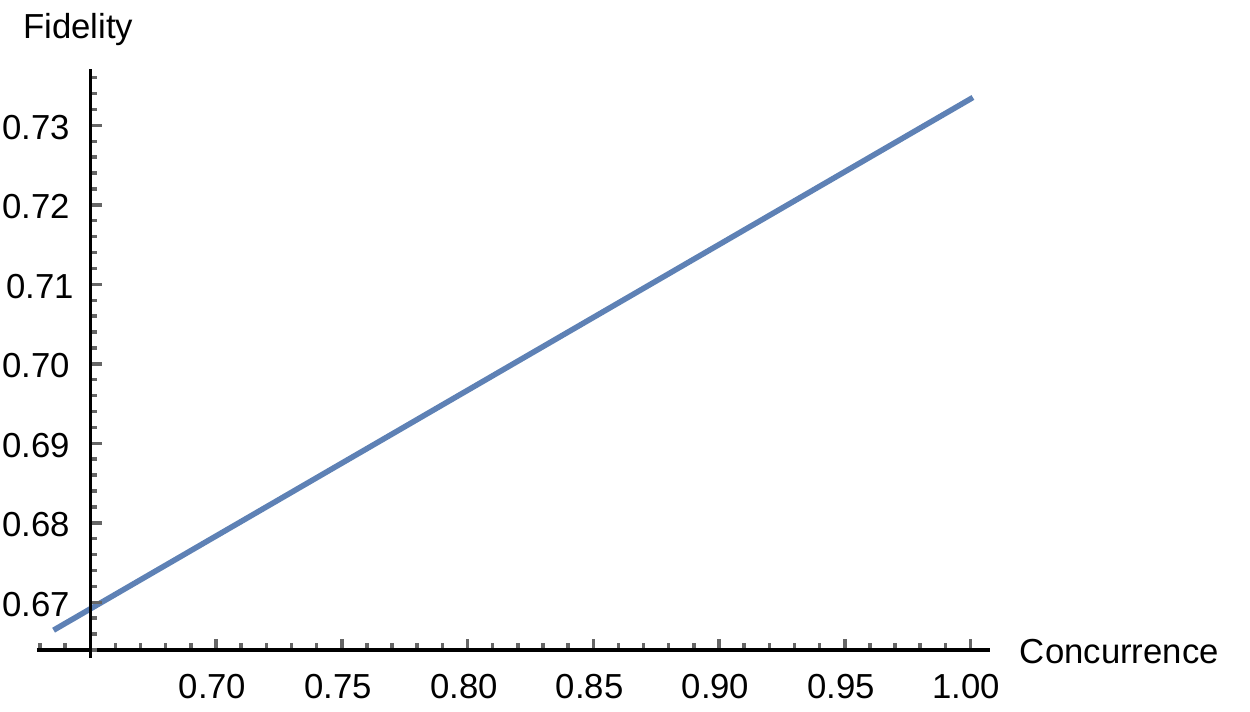}  
\caption{Teleportation fidelity is plotted against the concurrence of a pure entangled state when a particular type of noisy classical channel ($p_0=0.6$, $p_1=0.2$, $p_2=0.15$, and $p_3=0.05$) is employed. In the plot, it is clearly shown that the teleportation fidelity belongs to the non-classical range.}\label{fig1}
\end{center}
\end{figure}

\begin{figure}[h!]
\begin{center}
\includegraphics[width=.6\textwidth, height=0.28\textheight]{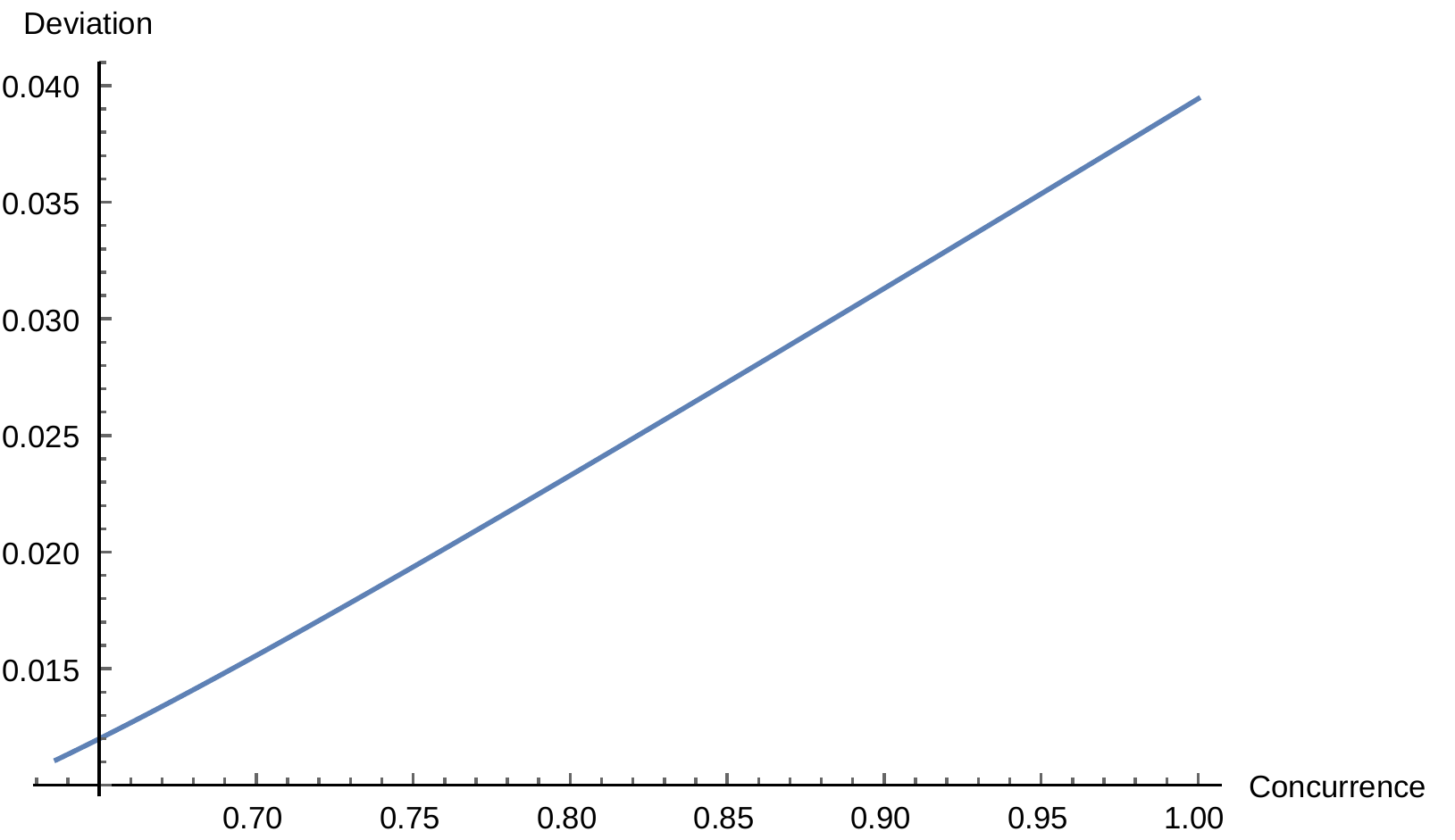}  
\caption{Fidelity deviation is plotted against the concurrence of a pure entangled state when a particular type of noisy classical channel ($p_0=0.6$, $p_1=0.2$, $p_2=0.15$, and $p_3=0.05$) is employed. In the plot, it is clearly shown that the fidelity deviation increases with the increase of entanglement in the resource state.}\label{fig2}
\end{center}
\end{figure}

{\it Pure entangled states:} We consider the state $\ket{\Phi}=a\ket{00}+b\ket{11}$, $a$ and $b$ are non-zero positive real numbers, such that $a^2+b^2=1$. For this state $|\mathbf{t}_{11}| = |\mathbf{t}_{22}| = 2ab$ and $|\mathbf{t}_{33}|=1$. So, $\mathbf{F}_\varrho$ and $\Delta_\varrho$, corresponding to the state $\ket{\Phi}$ under the Noise Model-I, are given by-
\begin{eqnarray}\label{eq33}
\mathbf{F}_{\varrho} &=& \frac{2}{3}(1+ab)-\frac{1}{3}\left[p_1+p_2+2ab(p_1+p_2+2p_3)\right], \nonumber\\
\Delta_{\varrho} &=& \frac{1}{\sqrt{30}}\left[(p_0+p_2-p_1-p_3)^24a^2b^2+(p_0+p_1-p_2-p_3)^24a^2b^2 \right.\\
&&~~~~~~~~~~+(p_0+p_3-p_1-p_2)^2-1/3((p_0+p_2-p_1-p_3)2ab \nonumber \\
&&~~~~~~~~~~\left.+(p_0+p_1-p_2-p_3)2ab+(p_0+p_3-p_1-p_2))^2\right]^{\frac{1}{2}}. \nonumber
\end{eqnarray}
In the above expressions if we put $a=b=\frac{1}{\sqrt{2}}$, assuming $\ket{\Phi}$ as a Bell state (i.e., $\ket{\Phi}$ = $\ket{\phi_0}$) then we get $\mathbf{F}_\varrho$ = $\frac{1}{3}(1+2p_0)$ and $\Delta_{\varrho}$ = $(2/3\sqrt{10})[(p_1-p_2)^2+(p_1-p_3)^2+(p_2-p_3)^2]^{1/2}$. So, for the Bell state, $\Delta_{\varrho}=0$ only when $p_1=p_2=p_3$. If we assume $p_1=p_2=p_3$, considering $a>b$ in the above expression of $\Delta_{\varrho}$, then we get $\Delta_{\varrho}$ = $(1/3\sqrt{5})(1-4p_1)(1-2ab)$, which is decreasing function of the entanglement (we use concurrence \cite{Wootters98} here as a measure of entanglement), contained by the state $\ket{\Phi}$, i.e., $2ab$. Nevertheless, if $p_0\neq p_1\neq p_2\neq p_3$, then it is possible to demonstrate specific scenarios where the fidelity deviation increases with the increment of entanglement, contained in the given pure state as resource. One such scenario is given as the following: We consider $p_0=0.6$, $p_1=0.2$, $p_2=0.15$, and $p_3=0.05$ and plot $\mathbf{F}_{\varrho}$, $\Delta_{\varrho}$ with respect to the concurrence of $\ket{\Phi}$, i.e., $2ab$. These plots are given in Fig.~(\ref{fig1}) and Fig.~(\ref{fig2}). From the figures, it is clear that the teleportation fidelity is non-classical when the concurrence of the state $\ket{\Phi}$ is greater than $0.64$. Furthermore, in this range, the deviation is increasing with the increment of the concurrence of $\ket{\Phi}$.

In the present protocol, if the classical communication is noiseless, then the protocol is known to be an optimal protocol \cite{Horodecki96, Badziag00}. This is in the sense that when the classical communication is noiseless, the average fidelity becomes maximum. However, in such an optimal protocol, pure entangled states are not dispersion-free. But these state might be dispersion-free when there is noise in the classical communication. We provide an explicit example where such thing is happening. 

We assume $a=\sqrt{0.9}$ and $b=\sqrt{0.1}$, thus, the concurrence of the state $\ket{\Phi}$ = $2ab=0.6$. To make this state dispersion-free, we search for a noisy classical channel which obeys the Noise Model-I. Using Eq.~(\ref{eq31}), we get $p_1=p_2$ as $|\mathbf{t}_{11}|=|\mathbf{t}_{22}|=0.6$. We also assume $p_1=p_2=0.15$. This implies that $p_3$ must be 0.017 to satisfy Eq.~(\ref{eq31}). Therefore, the state $\sqrt{0.9}\ket{00}+\sqrt{0.1}\ket{11}$ is dispersion-free in our protocol if the available noisy classical channel is characterized by the probabilities $p_1=p_2=0.15$, $p_3=0.017$, and $p_0=(1-0.317)=0.683$. It is also important to check if this state can lead to non-classical fidelity in our protocol when the available noisy classical channel is characterized by the above probabilities. Putting the values $|\mathbf{t}_{11}|=|\mathbf{t}_{22}|=0.6$, $|\mathbf{t}_{33}|=1$, $p_1=p_2=0.15$, $p_3=0.017$ in Eq.~(\ref{eq19}), we get $\mathbf{F}_\varrho=0.7$, which is greater than $2/3$ and thus, it is non-classical fidelity. This example indicates the fact that even if a resource state is provided which is not dispersion-free in an optimal protocol, it might be dispersion-free if a suitable noisy classical channel is employed without compromising the non-classical fidelity. 

It is also possible to consider mixed states which exhibit the above phenomenon. Such a state is given by- $\rho$ = $\frac{1}{4}(\mathbb{I}_4-\sum_{i=1}^3|\mathbf{t}_{ii}|\sigma_i\otimes\sigma_i)$, where $|\mathbf{t}_{11}|=|\mathbf{t}_{22}|=0.6$, and $|\mathbf{t}_{33}|=1$. This mixed state along with the above noisy classical channel exhibit the above phenomenon. 

To find the expressions of $\mathbf{F}_{\varrho}$ and $\Delta_{\varrho}$ for $\ket{\Phi}=a\ket{00}+b\ket{11}$, using Noise Model-II, we have to use the transformation rules: $p_0\rightarrow \eta\eta^\prime, p_1\rightarrow(1-\eta)\eta^\prime, p_2\rightarrow\eta(1-\eta^\prime), p_3\rightarrow(1-\eta)(1-\eta^\prime)$ and have to modify Eq.~(\ref{eq33}). The modified expressions are given by-
\begin{equation}\label{}
\begin{array}{l}
\mathbf{F}_{\varrho}=\frac{2}{3}(1+ab)-\frac{1}{3}\left[2ab(2-\eta-\eta^\prime)+(\eta+\eta^\prime-2\eta\eta^\prime)\right],\\
\Delta_{\varrho}=\frac{1}{3\sqrt{10}}[16a^2b^2(\eta-\eta^\prime)^2+(2\eta^\prime-1)^2(2ab-2\eta+1)^2+\\
~~~~~~~~(2\eta-1)^2(2ab-2\eta^\prime+1)^2].
\end{array}
\end{equation}
In the above expressions if we put $a=b=1/\sqrt{2}$, assuming $\ket{\Phi}$ as a Bell state (i.e., $\ket{\Phi}$ = $\ket{\phi_0}$) then we get $\mathbf{F}_\varrho$ = $\frac{1+2\eta\eta^\prime}{3}$ and $\Delta_{\varrho}$ = $(2/3\sqrt{5})[1-3\eta+3\eta^2+\eta^\prime(-3+7\eta-6\eta^2)+\eta^{\prime2}(3-6\eta+4\eta^2)]^{1/2}$. However, assuming $\ket{\Phi}$ as a nonmaximally entangled state, we can observe the variation of fidelity deviation with the concurrence of $\ket{\Phi}$. In particular, like Noise Model-I, here also we can demonstrate the increase of fidelity deviation with the increment of the concurrence, $2ab$. 

{\it Werner states:} We consider a family of Werner states \cite{Werner89}, given by- $\rho_W$ = $\epsilon\ket{\phi_0}\bra{\phi_0}+\frac{1}{4}(1-\epsilon)\mathbb{I}_4$, where $\ket{\phi_0}$ = $(1/\sqrt{2})(\ket{00}+\ket{11})$. The states $\rho_W$ are entangled when $1/3<\epsilon\leq1$ (we consider here only these values of $\epsilon$). For $\rho_W$, it is known that $|\mathbf{t}_{11}|$ = $|\mathbf{t}_{22}|$ = $|\mathbf{t}_{33}|$ = $\epsilon$. Using these values in Eq.~(\ref{eq19}), we get-
\begin{equation}\label{eq35}
\mathbf{F}_{\varrho} = \frac{1}{6}(3-\epsilon+4\epsilon p_0).
\end{equation}
It is consistent with the results of Ref.~\cite{Banik13}.

We first use Noise Model-I. Again, using Eq.~(\ref{eq29}), we get the expression of $\Delta_{\varrho}$ for $\rho_W$, which is given by-
\begin{equation}\label{eq36}
\begin{array}{c}
\Delta_{\varrho} = \frac{4(\frac{3\epsilon-1}{2})+2}{9\sqrt{10}}[(p_1-p_2)^2+(p_1-p_3)^2+(p_2-p_3)^2]^{1/2},
\end{array}
\end{equation}
where the term $\frac{3\epsilon-1}{2}$ is the concurrence of $\rho_W$. From the above expression, it is quite evident that for Noise Model-I, the fidelity deviation, corresponding to the state $\rho_W$, is zero only when $p_1=p_2=p_3$. On the other hand, for the fixed values of $p_1,p_2,p_3$, where $p_1\neq p_2\neq p_3$, the fidelity deviation, corresponding to the state $\rho_W$, is an increasing function of concurrence of the state. Using Noise Model-II also, one can get the expressions of teleportation fidelity and its deviation for $\rho_W$. To serve this purpose, one has to use the transformation rules: $p_0\rightarrow \eta\eta^\prime, p_1\rightarrow(1-\eta)\eta^\prime, p_2\rightarrow\eta(1-\eta^\prime), p_3\rightarrow(1-\eta)(1-\eta^\prime)$ in Eq.~(\ref{eq35}) and in Eq.~(\ref{eq36}). We mention here that when the classical communication is noiseless in the present protocol, pure states are not dispersion-free but the Werner states ($\rho_W$) are. This is as presented in Ref.~\cite{Ghosal20-1}. But when classical channels are noisy then we get a different result from that of Ref.~\cite{Ghosal20-1}. Therefore, one may expect that when the classical communication is noisy, Werner states may perform better compared to the pure states from fidelity deviation point of view. But we find here specific situations where the fidelity deviation increases with the increment of entanglement within the Werner states.

{\it Special cases:} From the expression, given in Eq.~(\ref{eq19}), it is expected that the maximum value of the teleportation fidelity is dependent on the given resource state as well as on the noisy classical channel. However, if the resource state and the noisy classical channel both are arbitrary, then it is difficult to provide exact conditions for which the fidelity is maximum. But we focus on two cases which are worth mentioning. (i) We assume that the resource state is given and it has the property: $|\mathbf{t}_{11}|$ = $|\mathbf{t}_{22}|$ = $|\mathbf{t}_{33}|$. In this case, if we consider the teleportation fidelity of Eq.~(\ref{eq19}), then the fidelity depends on the values of $|\mathbf{t}_{11}|$ and $p_0$. Notice that for the Noise Model-I, for a fixed value of $p_0$, it is possible to consider different noisy classical channels for which $p_1+p_2+p_3$ = $1-p_0$ (fixed) but the individual values of $p_1, p_2, p_3$ are different. We say this class of channels as $\Lambda_{p_0}$. Interestingly, the given resource state leads to same amount of teleportation fidelity in the present protocol for all the channels belonging to the class $\Lambda_{p_0}$ with a fixed $p_0$. In this sense, any state with the property $|\mathbf{t}_{11}|$ = $|\mathbf{t}_{22}|$ = $|\mathbf{t}_{33}|$, are not biased towards a particular channel of $\Lambda_{p_0}$ (with a fixed $p_0$). (ii) We assume that the noisy classical channel is given (under Noise Model-I) and it has the property: $p_1=p_2=p_3$. In this case, if we consider the teleportation fidelity of Eq.~(\ref{eq19}), then the fidelity depends on the value of the sum ($\sum_{i=1}^3|\mathbf{t}_{ii}|$) and on $p_1$. Now, for a fixed value of the sum ($\sum_{i=1}^3|\mathbf{t}_{ii}|$), we can have different states. We say this class of states with a fixed value of the sum ($\sum_{i=1}^3|\mathbf{t}_{ii}|$) as $\Lambda_{\varrho}$. Interestingly, the given channel leads to same amount of teleportation fidelity in the present protocol for all the states belonging to the class $\Lambda_{\varrho}$ with a fixed value of the sum ($\sum_{i=1}^3|\mathbf{t}_{ii}|$). In this sense, any channel with the property $p_1=p_2=p_3$, are not biased towards a particular state of $\Lambda_{\varrho}$ with a fixed value of the sum ($\sum_{i=1}^3|\mathbf{t}_{ii}|$). We also mention that when a resource state with the property: $|\mathbf{t}_{11}|$ = $|\mathbf{t}_{22}|$ = $|\mathbf{t}_{33}|$, is given or a noisy classical channel with the property: $p_1=p_2=p_3$ (under Noise Model-I), is given, the expression of Eq.~(\ref{eq19}), provides the maximum achievable fidelity within the present protocol. 
\vskip 0.1 in

{\bf Minimum classical communication cost:} From Eq.~(\ref{eq21}), we find the condition for non-classical fidelity under Noise Model-I. However, here our motivation is to obtain the minimum amount of classical information which is required to communicate from Alice's side to Bob, such that the condition for non-classical fidelity, given in Eq.~(\ref{eq21}), is satisfied for any two-qubit resource state. We say this minimum amount of classical information as minimum classical communication cost (within our protocol). The quantification of classical information is given in Eq.~(\ref{eq4}), $\mathbf{I}$ = $2+\sum_{i=0}^3p_i\log_2(p_i)$, where $\sum_{i=0}^3p_i$ = 1. Now, we formulate the problem of minimum classical communication cost in the following way: 
\begin{equation}\label{eq40}
\begin{array}{l}
\mbox{minimize}~~~~2+\sum_{i=0}^3p_i\log_2(p_i)\\[1 ex]
\mbox{subject to}~~\sum_{i=1}^3|\mathbf{t}_{ii}|>(1+2f_{noise}),\\[1 ex]
~~~~~~~~~~~~~~~\sum_{i=0}^3 p_i=1.
\end{array}
\end{equation}

To get the solution of the above problem, we first consider the constraint equations with equality. Then, we can apply Lagrange's method of undetermined multipliers. It is difficult to get the exact analytical solutions because we get transcendental equations. But using this method one can obtain some  stationary conditions for which the minimization can be obtained. The conditions are 

\begin{equation}\label{eq41}
p_0=\frac{1}{2(|t_{ii}|+|t_{kk}|)} \left(1+|t_{ii}|-|t_{jj}|+|t_{kk}|+2p_2(|t_{jj}|-|t_{ii}|)+2p_3(|t_{jj}|-|t_{kk}|)\right)
\end{equation}
and 
\begin{equation}\label{eq42}
p_1=\frac{1}{2(|t_{ii}|+|t_{kk}|)} \left(-1+|t_{ii}|+|t_{jj}|+|t_{kk}|-2p_2(|t_{jj}|+|t_{kk}|)-2p_3(|t_{ii}|+|t_{jj}|) \right)
\end{equation}

So, if Eq.~(\ref{eq41}) and Eq.~(\ref{eq42}) hold for some values of $p_0, p_1, p_2, p_3$ where $p_i\geq0$ and $\sum_{i=0}^3p_i=1$, then substituting these values into Eq.~(\ref{eq40}) one can get the minimum classical communication cost to accomplish the non-classical teleportation fidelity.

Similarly, we can find the minimum classical communication cost conditions for non-classical fidelity under Noise Model-II where Eq.~(\ref{eq26}) is satisfied for any two-qubit resource state. Now, for Noise Model-II, the quantification of classical information is given in Eq.~(\ref{eq6}), $I_l=1-H(l)$ where $H(l)=-l\log_2l-(1-l)\log_2(1-l)$.  So, the problem of minimum classical communication cost is
\begin{equation}\label{eq43}
\begin{array}{l}
\mbox{minimize}~~~~2+\eta\log_2\eta+(1-\eta)\log_2(1-\eta)+\eta^{\prime}\log_2\eta^{\prime}+(1-\eta^{\prime})\log_2(1-\eta^{\prime})\\[1 ex]
\mbox{subject to}~~\sum_{i=1}^3|\mathbf{t}_{ii}|>(1+2f_{noise}^{\prime}),\\[1 ex]
~~~~~~~~~~~~~~~\frac{1}{2}\leq\eta\leq1, \frac{1}{2}\leq\eta^{\prime}\leq1.
\end{array}
\end{equation}
As we want to calculate the minimum classical communication cost first we consider the constraint equations with equality sign.  We use here Lagrangian multiplier method to get solutions. Again, it is difficult to get exact analytical solutions because we get transcendental equations. But we get only stationary conditions and these are

\begin{eqnarray}\label{eq44}
\frac{\log_2\eta-\log_2(1-\eta)}{(|t_{ii}|-(1-2\eta^{\prime})|t_{kk}|)}=\frac{\log_2\eta^{\prime}-\log_2(1-\eta^{\prime})}{(|t_{jj}|-(1-2\eta)|t_{kk}|)}
\end{eqnarray}
and
\begin{equation}\label{eq45}
\sum_{i=1}^3|t_{ii}|-1=2\bigg(|t_{ii}|(1-\eta)+|t_{jj}|(1-\eta^{\prime})+|t_{kk}|(\eta+\eta^{\prime}-2\eta\eta^{\prime})\bigg)
\end{equation}
If the above equations get satisfied by the variables $\eta$ and $\eta^{\prime}$ within the ranges $\frac{1}{2}\leq\eta\leq1$, $\frac{1}{2}\leq\eta^{\prime}\leq1$ then one can find minimum classical communication cost by putting $\eta$, $\eta^{\prime}$ values into the Eq.~(\ref{eq6}) after knowing the exact values of all $|t_{ii}|, i\in\{1,2,3\}$.

We see that for both the minimization problems, it is difficult to get an exact solution of the problems. However, for a given state, it may possible to solve the above. For example, for a pure state $\ket{\Phi}$, we can put $|\mathbf{t}_{11}|$ = $|\mathbf{t}_{22}|$ = $2ab$ and $|\mathbf{t}_{33}|$ = 1 in the above equations, where $\ket{\Phi}$ = $a\ket{00}+b\ket{11}$, $a,b$ are nonzero positive numbers such that $a^2+b^2=1$. Thus, we can have expressions of $p_i$ in terms of the entanglement of $\ket{\Phi}$ (i.e., $2ab$) $\forall i=0,\dots3$. Putting these values of $p_i$, in Eq.~(\ref{eq4}), one can get the minimum classical communication cost. Using Noise Model-II also, one can follow similar type of procedure to get minimum classical communication cost.
\vskip 0.1 in

{\bf Optimality of teleportation fidelity:} If the classical communication is noiseless then, it is already known what is the optimal teleportation fidelity \cite{Badziag00}. Nevertheless, we are now going to prove that even if there is noise (particularly, the type of noises that we have chosen) in the classical communication, the teleportation fidelity which we have derived is optimal and it can be achieved through the present protocol.

We assume that classical communication is restricted (not more than two cbits of information) in our protocol. Therefore, we do not allow any classical error correcting protocol. This is because when classical communication is limited, it is not possible to apply classical error correcting code to get the right information from noisy classical channel(s). Actually, to get the right information from every classical bit, one has to send it many times through noisy classical channel(s). 

Remember that to maximize the teleportation fidelity, it is necessary to maximize the fidelity of the quantum resource state (fully entangled fraction). In our protocol we have already maximized the fully entangled fraction of quantum resource state as we have started with a resource state in canonical form (see \cite{Horodecki96, Badziag00, Albeverio02} in this regard). 

Next, to get optimal teleportation fidelity, we need to find bob's optimal unitary correction. We can now summarize the steps to obtain optimal teleportation fidelity. (i) First, Alice and Bob share an entangled state in canonical form as resource. This state is $\varrho$. (ii) Alice does the Bell basis measurement on her qubits and sends the measurement outcome to Bob through a classical channel which is noisy. (iii) As the classical communication is restricted (not more than two cbits), we do not allow any classical error correcting code to get the right information. (iv) Next, Bob will apply optimal unitary operator to his qubit to get maximal teleportation fidelity.

The final state after following the teleportation protocol with noiseless classical communication is given by-
\begin{equation}\label{eq1a}
\varrho_{avg}=\sum_{k=0}^3(\mathbb{I}_4\otimes V_{k})\frac{1}{s_k}\mbox{Tr}_{A}\left[(|\phi_k\rangle\langle\phi_k|\otimes\mathbb{I}_2)(|\psi\rangle\langle\psi|\otimes\varrho)(|\phi_k\rangle\langle\phi_k|\otimes\mathbb{I}_2)\right](\mathbb{I}_4\otimes V_{k}^{\dagger}),
\end{equation}
where $\{|\phi_{k}\rangle\}$ are the Bell states. $|\psi\rangle$ is the unknown state to be teleported and $V_k$ are the Bob's unitary operators, for more details see (\ref{eq9}). Bob chooses the unitary operator from the set $V_k\in\{\varsigma_0,\varsigma_1,\varsigma_2,\varsigma_3\}$. Alice's measurement is a four outcome measurement and thus, Bob assigns one unitary for each measurement outcome from $\{\varsigma_0, \varsigma_1,\varsigma_2,\varsigma_3\}$ set.

When we follow Noise Model-I, Bob applies probabilistic unitary operators given in Table \ref{tab2}. So, in this case the average output state can be written as: 
\begin{equation}\label{eq2a}
\varrho_{avg}=\sum_{i=0}^3\sum_{k=0}^3p_i(\mathbb{I}_4\otimes V_{k i})\frac{1}{s_k}\mbox{Tr}_{A}\left[(|\phi_k\rangle\langle\phi_k|\otimes\mathbb{I}_2)(|\psi\rangle\langle\psi|\otimes\varrho)(|\phi_k\rangle\langle\phi_k|\otimes\mathbb{I}_2)\right](\mathbb{I}_4\otimes V_{k i}^{\dagger}),
\end{equation}
where the probabilities $p_i$ are defined earlier. Notice that $p_0$ is responsible for appropriate unitary rotation and $p_1$, $p_2$, $p_3$ introduce errors to Bob’s choice of unitary after receiving the information of Alice’s
measurement outcome. So, basically here one can get the following relations:
\begin{center}
$\varsigma_0=V_{00}=V_{13}=V_{22}=V_{31}$,\\[1 ex]
$\varsigma_1=V_{10}=V_{03}=V_{21}=V_{32}$,\\[1 ex]
$\varsigma_2=V_{20}=V_{11}=V_{02}=V_{33}$,\\[1 ex]
$\varsigma_3=V_{30}=V_{01}=V_{12}=V_{23}$.
\end{center}
In this way one can get back (\ref{eq12}). Note that after Alice's measurement and before Bob's unitary operation the state corresponding to each outcome is given in (\ref{eq10}). We now head to provide the proof of optimality in a greater details. 

Lets assume that $\mathcal{O}_0$, 
$\mathcal{O}_1$, $\mathcal{O}_2$, and $\mathcal{O}_3$ are optimal orthogonal representations ($\mathcal{O}_k$s are rotations in real space $\mathbb{R}^3$. $\mathcal{O}_k$ can be determined uniquely as the group of rotations O(3) is a homomorphic image of U(2) group) of unitary operators which are applied by Bob and this maximizes the teleportation fidelity. From (\ref{eq12}), we now consider the following term: $s_0\varsigma_0\varrho_0\varsigma_0+s_1\varsigma_1\varrho_1\varsigma_1 + s_2\varsigma_2\varrho_2\varsigma_2 + s_3\varsigma_3\varrho_3\varsigma_3$. Using the concept $\varsigma_k(\mbox{\bf{n}}\bm{\cdot\sigma})\varsigma_k=(\mathcal{O}^{\dag}_k\mbox{\bf{n}})\bm{\cdot\sigma}$, the aforesaid term can be rewritten and it is given in (\ref{eq13}). Similar terms of (\ref{eq12}) can also be rewritten in the same way. After rewriting these terms, we can sum them up and get the following:
\begin{eqnarray}\label{eq3a}
\varrho_{avg}=\frac{1}{2}\mathbb{I}_{2}+\frac{1}{8}\sum_{k=0}^{3}(\mathcal{O}_k^{\dagger}\mbox{\bf{s}})\bm{\cdot\sigma} +\frac{1}{8}(\mbox{\bf{X a}})\bm{\cdot\sigma},
\end{eqnarray}
where {\bf X} is given by-
\begin{equation}\label{eq4a}
\begin{array}{r}
\mbox{\bf X} = p_0(\mathcal{O}_0^{\dagger}\mathbb{T}^{\dagger}\mathbb{T}_0+\mathcal{O}_1^{\dagger}\mathbb{T}^{\dagger}\mathbb{T}_1+\mathcal{O}_2^{\dagger}\mathbb{T}^{\dagger}\mathbb{T}_2+\mathcal{O}_3^{\dagger}\mathbb{T}^{\dagger}\mathbb{T}_3)+\\[0.5 ex] p_1(\mathcal{O}_3^{\dagger}\mathbb{T}^{\dagger}\mathbb{T}_0+\mathcal{O}_2^{\dagger}\mathbb{T}^{\dagger}\mathbb{T}_1+\mathcal{O}_1^{\dagger}\mathbb{T}^{\dagger}\mathbb{T}_2+\mathcal{O}_0^{\dagger}\mathbb{T}^{\dagger}\mathbb{T}_3)+\\[0.5 ex]
p_2(\mathcal{O}_2^{\dagger}\mathbb{T}^{\dagger}\mathbb{T}_0+\mathcal{O}_3^{\dagger}\mathbb{T}^{\dagger}\mathbb{T}_1+\mathcal{O}_0^{\dagger}\mathbb{T}^{\dagger}\mathbb{T}_2+\mathcal{O}_1^{\dagger}\mathbb{T}^{\dagger}\mathbb{T}_3)+\\[0.5 ex]
p_3(\mathcal{O}_1^{\dagger}\mathbb{T}^{\dagger}\mathbb{T}_0+\mathcal{O}_0^{\dagger}\mathbb{T}^{\dagger}\mathbb{T}_1+\mathcal{O}_3^{\dagger}\mathbb{T}^{\dagger}\mathbb{T}_2+\mathcal{O}_2^{\dagger}\mathbb{T}^{\dagger}\mathbb{T}_3).~~
\end{array}
\end{equation}
Notice that in the above $\varrho_{avg}$ is different from that of (\ref{eq14}). This is because we have not assumed here $\mathcal{O}_k^\dagger=-\mathbb{T}_k$. So, the expression of average fidelity is going to be different and it is given by-
\begin{equation}\label{eq5a}
\mbox{\bf{F}}_{\varrho}=\frac{1}{2}+\frac{1}{24}\mbox{Tr(\bf{X})}.
\end{equation}
We get the above expression by omitting the term which do not contribute to the average fidelity (when averaging over all input states) and using Schur's Lemma. From the above, it is clear that if we want to get the optimal teleportation fidelity then we have to maximize the term $\mbox{Tr(\bf{X})}$ and this term is totally dependent upon the Bob's unitary corrections. Recall that $\mathbb{T}$ corresponds to $\varrho$. When $\varrho$ is entangled, i.e., det$\mathbb{T}\leq0$, we can have $\mathbb{T}$ = diag($-|\mbox{\bf{t}}_{11}|, -|\mbox{\bf{t}}_{22}|, -|\mbox{\bf{t}}_{33}|$). This is already defined earlier. Furthermore, $\forall k, \mathbb{T}_k$ is also defined earlier, they belong to $\ket{\phi_k}$, the Bell states. Next, we are going to find what is the form of $\mathcal{O}_k^{\dagger}$ which contributes a maximum value to (\ref{eq5a}). To do this, we rewrite {\bf X} as:
\begin{eqnarray}\label{6a}
\mbox{\bf{X}} =  \mathcal{O}_0^{\dagger}\mathbb{T}^{\dagger}(p_0\mathbb{T}_0+p_1\mathbb{T}_3+p_2\mathbb{T}_2+p_3\mathbb{T}_1)\nonumber~\\
+\mathcal{O}_1^{\dagger}\mathbb{T}^{\dagger}(p_0\mathbb{T}_1+p_1\mathbb{T}_2+p_2\mathbb{T}_3+p_3\mathbb{T}_0)\nonumber~\\
+\mathcal{O}_2^{\dagger}\mathbb{T}^{\dagger}(p_0\mathbb{T}_2+p_1\mathbb{T}_1+p_2\mathbb{T}_0+p_3\mathbb{T}_3)\nonumber~\\
+\mathcal{O}_3^{\dagger}\mathbb{T}^{\dagger}(p_0\mathbb{T}_3+p_1\mathbb{T}_0+p_2\mathbb{T}_1+p_3\mathbb{T}_2).
\end{eqnarray}
Let us first maximize the first term of the above equation because the maximum value of Tr({\bf X}) is dependent on the individual maximization of these terms. However, the first term can be rewritten as $\mathcal{O}_0^{\dagger}\mathcal{X}_1$, where $\mathcal{X}_1=\mathbb{T}^{\dagger}(p_0\mathbb{T}_0+p_1\mathbb{T}_3+p_2\mathbb{T}_2+p_3\mathbb{T}_1)$ is a $3\times3$ diagonal matrix and $\mathcal{O}_0^{\dagger}$ is a $3\times3$ matrix. For a given resource state and a classical channel $|\mbox{\bf{t}}_{11}|,|\mbox{\bf{t}}_{22}|,|\mbox{\bf{t}}_{33}|,p_0,p_1,p_2,p_3$ are fixed. Ultimately, to maximize $\mbox{Tr}(\mathcal{O}_0^{\dagger}\mathcal{X}_1)$, we have to maximize diagonal elements of the matrix, $\mathcal{O}_0^{\dagger}$ (remembering $\det(\mathcal{O}_0) = +1$) because only its diagonal elements contribute to the trace value of $\mbox{Tr}(\mathcal{O}_0^{\dagger}\mathcal{X}_1)$. Let us assume that $\mathcal{O}_0^{\dagger}$ has diagonal elements $x_1$, $y_1$ and $z_1$. So, $\mbox{Tr}(\mathcal{O}_0^{\dagger}\mathcal{X}_1)$ can be rewritten as the following:
\begin{eqnarray}\label{eq7a}
\mbox{Tr}(\mathcal{O}_0^{\dagger}\mathcal{X}_1) & = & p_0(-x_1|\mbox{\bf{t}}_{11}|+y_1|\mbox{\bf{t}}_{22}|-z_1|\mbox{\bf{t}}_{33}|) + p_1(x_1|\mbox{\bf{t}}_{11}|+y_1|\mbox{\bf{t}}_{22}|+z_1|\mbox{\bf{t}}_{33}|)\nonumber\\
&& + p_2(-x_1|\mbox{\bf{t}}_{11}|-y_1|\mbox{\bf{t}}_{22}|+z_1t_3) + p_3(x_1|\mbox{\bf{t}}_{11}|-y_1|\mbox{\bf{t}}_{22}|-z_1|\mbox{\bf{t}}_{33}|).
\end{eqnarray}
In general, for a qubit teleportation, the unitary operators which belong to SU(2) group, forms an orthogonal representation of the rotation SO(3) in real space ($\mathbb{R}^3$). In fact, $\mathcal{O}_k$ is a proper rotation and det($\mathcal{O}_k$) = +1. These rotations form a group where it satisfies $\mathcal{O}_k^T \mathcal{O}_k = \mathbb{I}_3$. We assume $\det(\mathcal{O}_k)=\lambda_1\lambda_2\lambda_3$, where $\lambda_1$, $\lambda_2$, $\lambda_3$ are eigenvalues of $\mathcal{O}_k$. So, for $\mathcal{O}_k$, it can be shown that there is always an eigenvalue $+1$. In terms of Euler angles one can find for a general rotation about an arbitrary axis \cite{MR2276052,book}.
\begin{equation}\label{eq8a}
\mathcal{O}_k=\begin{bmatrix}
\cos\phi\cos\psi-\cos\theta\sin\phi\sin\psi & -\cos\phi\sin\psi-\cos\theta\sin\phi\cos\psi & \sin\phi\sin\theta\\
\sin\phi\cos\psi+\cos\theta\cos\phi\sin\psi & -\sin\phi\sin\psi+\cos\theta\cos\phi\cos\psi & -\cos\phi\sin\theta\\
\sin\psi\sin\theta & \cos\psi\sin\theta & \cos\theta
\end{bmatrix} \mbox{(say)}.
\end{equation}    
From the general arbitrary rotation of $\mathcal{O}_k$, given above, it is observed that diagonal maximum values can be found either when two of the diagonal elements are $-1$ and another one will be $+1$ or when all the diagonal elements are $+1$ satisfying $\det(\mathcal{O}_k)=+1$. Then, all the off-diagonal elements will be zero. 

From (\ref{eq7a}), $\mbox{Tr}(\mathcal{O}_0^{\dagger}\mathcal{X}_1)$ is a linear function of $x_1$, $y_1$, and $z_1$ for given values of $|\mbox{\bf{t}}_{11}|,|\mbox{\bf{t}}_{22}|,|\mbox{\bf{t}}_{33}|, p_0, p_1, p_2, p_3$. We note that $\sum_i p_i = 1$, $0\leq p_i \leq 1$ $\forall i$, $|\mbox{\bf{t}}_{11}|\geq |\mbox{\bf{t}}_{22}|\geq |\mbox{\bf{t}}_{33}|$ and $0\leq |\mbox{\bf{t}}_{11}|,|\mbox{\bf{t}}_{22}|,|\mbox{\bf{t}}_{33}|\leq1$. Therefore, if $p_0>p_1,p_2,p_3$ then, to maximize $\mbox{Tr}(\mathcal{O}_0^{\dagger}\mathcal{X}_1)$, one should choose $x_1=-1$, $y_1=1$ and $z_1=-1$. So, we now can write $\mathcal{O}_0^{\dagger}$ as:
\begin{equation}\label{eq9a}
\mathcal{O}_0^{\dagger} = 
\begin{bmatrix}
-1 & 0 &  0\\
 0 & 1 &  0\\
 0 & 0 & -1
\end{bmatrix} = - 
\begin{bmatrix}
1 &  0 & 0\\
0 & -1 & 0\\
0 &  0 & 1
\end{bmatrix} = -
\mathbb{T}_0
\end{equation}
which is noting but $\sigma_2$ [a Pauli's unitary matrix in SU(2)], represented in SO(3). Note that the above can also be found using Karush-Kuhn-Tucker
(KKT) method. Following similar arguments, it is possible to show $\mathcal{O}_k^{\dagger}$ = $-\mathbb{T}_k$, for $k=1,2,3$. They are noting but $\sigma_1$, $\sigma_3$, and $\sigma_0$ respectively [Pauli's  unitary matrices in SU(2)], represented in SO(3). These are when $p_0>p_1,p_2,p_3$. We now summarize this case in the following. 
\begin{table}[h]
\begin{center}
\caption{Strategy for unitary corrections when $p_0>p_1,p_2,p_3$}\label{strategy1}
\begin{tabular}{|c|c|c|}
\hline 
States of Alice's measurement & Alice's classical message to Bob & Application of unitary operator by Bob\\
\hline\hline
$|\phi_0\rangle$ & 00 & $\varsigma_0=\sigma_2$\\
\hline
$|\phi_1\rangle$ & 11 & $\varsigma_1=\sigma_1$\\
\hline
$|\phi_2\rangle$ & 01 & $\varsigma_2=\sigma_3$\\
\hline
$|\phi_3\rangle$ & 10 & $\varsigma_3=\sigma_0$\\
\hline
\end{tabular}
\end{center}
\end{table}
For this strategy, the optimal teleportation fidelity becomes: 
\begin{equation}\label{eq10a}
\mbox{\bf{F}}_{\varrho}=\frac{1}{2}\left(1+\frac{1}{3}\sum_{i=1}^3|\mbox{\bf{t}}_{ii}|\right)-\frac{1}{3}\Bigg(p_1(|\mbox{\bf{t}}_{11}|+|\mbox{\bf{t}}_{33}|)+p_2(|\mbox{\bf{t}}_{22}|+|\mbox{\bf{t}}_{33}|)+p_3(|\mbox{\bf{t}}_{11}|+|\mbox{\bf{t}}_{22}|)\Bigg).
\end{equation}

Note that there are steps of obtaining the above expression. (i) From (\ref{eq9a}) the eigenvalues of $\mathcal{O}_0^\dagger$, can be put in (\ref{eq7a}). Then, we find the maximum value of Tr$(\mathcal{O}_0^\dagger\mathcal{X}_1)$. (ii) Following similar technique, we have to find out the maximum values of Tr$(\mathcal{O}_1^\dagger\mathcal{X}_2)$, Tr$(\mathcal{O}_2^\dagger\mathcal{X}_3)$, and Tr$(\mathcal{O}_3^\dagger\mathcal{X}_4)$, where $\mathcal{X}_2=\mathbb{T}^{\dagger}(p_0\mathbb{T}_1+p_1\mathbb{T}_2+p_2\mathbb{T}_3+p_3\mathbb{T}_0)$, $\mathcal{X}_3=\mathbb{T}^{\dagger}(p_0\mathbb{T}_2+p_1\mathbb{T}_1+p_2\mathbb{T}_0+p_3\mathbb{T}_3)$, and $\mathcal{X}_4=\mathbb{T}^{\dagger}(p_0\mathbb{T}_3+p_1\mathbb{T}_0+p_2\mathbb{T}_1+p_3\mathbb{T}_2)$. (iii) Actually, one can check the eigenvalues of $\mathcal{O}_1^\dagger$ to maximize Tr$(\mathcal{O}_1^\dagger\mathcal{X}_2)$. They are given by $1, -1, -1$. The eigenvalues of $\mathcal{O}_2^\dagger$ to maximize Tr$(\mathcal{O}_2^\dagger\mathcal{X}_3)$ are given by $-1, -1, 1$, and the eigenvalues of $\mathcal{O}_3^\dagger$ to maximize Tr$(\mathcal{O}_3^\dagger\mathcal{X}_4)$ are given by $1, 1, 1$. (iv) Then, we put these values of Tr$(\mathcal{O}_0^\dagger\mathcal{X}_1)$, 
Tr$(\mathcal{O}_1^\dagger\mathcal{X}_2)$,
Tr$(\mathcal{O}_2^\dagger\mathcal{X}_3)$, and Tr$(\mathcal{O}_3^\dagger\mathcal{X})$ in (\ref{eq5a}) to obtain the expression of (\ref{eq10a}).

For other three cases, when $p_1>p_0,p_2,p_3$ or $p_2>p_0,p_1,p_3$ or $p_3>p_0,p_1,p_2$, there will be change in strategy of unitary corrections and corresponding expression of average fidelity will also be changed. These can be obtained following the above technique. We now provide those summaries one by one. 

When $p_1>p_0,p_2,p_3$, we have to do the same process to find the $\{\mathcal{O}_k\}$s as we have done in the previous case. In this case, to maximize Tr({\bf X}), we have to choose $\mathcal{O}_0^{\dagger}=-\mathbb{T}_3$, $\mathcal{O}_1^{\dagger}=-\mathbb{T}_2$, $\mathcal{O}_2^{\dagger}=-\mathbb{T}_1$, and $\mathcal{O}_3^{\dagger}=-\mathbb{T}_0$. Then we have the following. 
\begin{table}[h!]
\begin{center}
\caption{Strategy for unitary corrections when $p_1>p_0,p_2,p_3$}\label{strategy2}
\begin{tabular}{|c|c|c|}
\hline 
States of Alice's measurement & Alice's classical message to Bob & Application of unitary operator by Bob\\
\hline\hline
$|\phi_0\rangle$ & 00 & $\varsigma_0=\sigma_0$\\
\hline
$|\phi_1\rangle$ & 11 & $\varsigma_1=\sigma_3$\\
\hline
$|\phi_2\rangle$ & 01 & $\varsigma_2=\sigma_1$\\
\hline
$|\phi_3\rangle$ & 10 & $\varsigma_3=\sigma_2$\\
\hline
\end{tabular}
\end{center}
\end{table}
In this strategy, the optimal teleportation fidelity is given by-
\begin{equation}\label{eq11a}
\mbox{\bf{F}}_{\varrho}=\frac{1}{2}\left(1+\frac{1}{3}\sum_{i=1}^3|\mbox{\bf{t}}_{ii}|\right)-\frac{1}{3}\Bigg(p_0(|\mbox{\bf{t}}_{11}|+|\mbox{\bf{t}}_{33}|)+p_2(|\mbox{\bf{t}}_{11}|+|\mbox{\bf{t}}_{22}|)+p_3(|\mbox{\bf{t}}_{22}|+|\mbox{\bf{t}}_{33}|)\Bigg).
\end{equation}

When $p_2>p_0,p_1,p_3$, we have to do the same process to find the $\{\mathcal{O}_k\}$s as we have done previously. In this case, to maximize Tr({\bf X}), we have to choose $\mathcal{O}_0^{\dagger}=-\mathbb{T}_2$, $\mathcal{O}_1^{\dagger}=-\mathbb{T}_3$, $\mathcal{O}_2^{\dagger}=-\mathbb{T}_0$, and $\mathcal{O}_3^{\dagger}=-\mathbb{T}_1$. Then we have the following.
\begin{table}[h]
\begin{center}
\caption{Strategy for unitary corrections when $p_2>p_0,p_1,p_3$}\label{strategy3}
\begin{tabular}{|c|c|c|}
\hline 
States of Alice's measurement & Alice's classical message to Bob & Application of unitary operator by Bob\\
\hline\hline
$|\phi_0\rangle$ & 00 & $\varsigma_0=\sigma_3$\\
\hline
$|\phi_1\rangle$ & 11 & $\varsigma_1=\sigma_0$\\
\hline
$|\phi_2\rangle$ & 01 & $\varsigma_2=\sigma_2$\\
\hline
$|\phi_3\rangle$ & 10 & $\varsigma_3=\sigma_1$\\
\hline
\end{tabular}
\end{center}
\end{table}
In this strategy, the optimal teleportation fidelity is given by-
\begin{equation}\label{eq12a}
\mbox{\bf{F}}_{\varrho}=\frac{1}{2}\left(1+\frac{1}{3}\sum_{i=1}^3|\mbox{\bf{t}}_{ii}|\right)-\frac{1}{3}\Bigg(p_0(|\mbox{\bf{t}}_{22}|+|\mbox{\bf{t}}_{33}|)+p_1(|\mbox{\bf{t}}_{11}|+|\mbox{\bf{t}}_{22}|)+p_3(|\mbox{\bf{t}}_{11}|+|\mbox{\bf{t}}_{33}|)\Bigg).
\end{equation}

When $p_3>p_0,p_1,p_2$, we have to do the same process to find the $\{\mathcal{O}_k\}$s as we have done previously. In this case, to maximize Tr({\bf X}), we have to choose $\mathcal{O}_0^{\dagger}=-\mathbb{T}_1$, $\mathcal{O}_1^{\dagger}=-\mathbb{T}_0$, $\mathcal{O}_2^{\dagger}=-\mathbb{T}_3$, and $\mathcal{O}_3^{\dagger}=-\mathbb{T}_2$. Then we have the following.
\begin{table}[h]
\begin{center}
\caption{Strategy for unitary corrections when $p_3>p_0,p_1,p_2$}\label{strategy4}
\begin{tabular}{|c|c|c|}
\hline 
States of Alice's measurement & Alice's classical message to Bob & Application of unitary operator by Bob\\
\hline\hline
$|\phi_0\rangle$ & 00 & $\varsigma_0=\sigma_1$\\
\hline
$|\phi_1\rangle$ & 11 & $\varsigma_1=\sigma_2$\\
\hline
$|\phi_2\rangle$ & 01 & $\varsigma_2=\sigma_0$\\
\hline
$|\phi_3\rangle$ & 10 & $\varsigma_3=\sigma_3$\\
\hline
\end{tabular}
\end{center}
\end{table}
In this strategy, the optimal teleportation fidelity is given by-
\begin{equation}\label{eq13a}
\mbox{\bf{F}}_{\varrho}=\frac{1}{2}\left(1+\frac{1}{3}\sum_{i=1}^3|\mbox{\bf{t}}_{ii}|\right)-\frac{1}{3}\Bigg(p_0(|\mbox{\bf{t}}_{11}|+|\mbox{\bf{t}}_{22}|)+p_1(|\mbox{\bf{t}}_{22}|+|\mbox{\bf{t}}_{33}|)+p_2(|\mbox{\bf{t}}_{11}|+|\mbox{\bf{t}}_{33}|)\Bigg).
\end{equation}

In this way we complete the proof of optimal teleportation fidelity for Noise Model-I. Now, to do the same for Noise Model-II, we just have to change the values of $p_i$ $\forall i = 0,1,2,3$.

\section{Conclusion and open problems}\label{sec4}
In this work, we have explored the combined effects of noisy resource state along with noisy classical communication on teleportation fidelity and its deviation. For this purpose, we have considered a teleportation protocol where an arbitrary two-qubit state in canonical form is given as resource along with noisy classical communication (Alice communicates not more than two cbits of information to Bob) to teleport an unknown qubit.  

Under the present teleportation protocol, we have derived the exact formulae of teleportation fidelity and its deviation. We prove the present teleportation fidelity is optimal, given the conditions, i.e., limited classical communication (not more than two cbits) through noisy channel(s). Furthermore, we have studied the conditions for non-classical fidelity and dispersion-free teleportation within this protocol. If the classical communication is noiseless then there are some resource states (specially, some noisy resource states) which can lead to zero fidelity deviation in the present teleportation protocol. However, we have shown that such resource states may not lead to zero fidelity deviation when the classical communication is noisy in the same protocol. Clearly, if all types of imperfections are there in a teleportation protocol, then in general, it is a difficult problem to find the resource states which can lead to both non-classical fidelity and zero fidelity deviation. So, we have to put more efforts to find a comprehensive picture of the types of imperfections which may occur in a teleportation protocol. 

One can argue that zero fidelity deviation for any two-qubit resource state is achievable via twirling protocols. Such a protocol is given as the following. Before teleportation, a random single qubit unitary operator from SU(2) according to the Haar measure can be applied by Alice to the input state and the inverse unitary operator can be applied by Bob to the output state after the teleportation is completed. In the whole process average teleportation fidelity will remain same. Alice and Bob can also follow a different protocol and there they can make the fidelity deviation zero by using random twirl, i.e., they can apply random unitary operators to transform the resource state into Werner form. For both the protocols, it requires high amount of classical communication (see also the discussion section of Ref.~\cite{Ghosal20-1}). Because the identities of the unitary operators must be communicated in each step. However, these protocols can be implemented in practical scenarios, but the accuracy might be limited as it depends on the implementations of the unitary operators.

In our analysis we have considered the combined effects of classical and quantum noises. In the above paragraph, we have discussed that one can get fidelity deviation zero by using twirling protocols but one has to give extra amount of classical communication which is necessary to disclose the identity of the unitary operation in each step. However, we are restricting ourselves to the limited classical communication cost along with noise in classical channels. So, our results are important as we are getting zero fidelity deviation in limited and noisy classical communication scenario. One may also argue that there are several efficient classical error correcting protocols, available to ensure noiseless classical communication. However, we have tried to highlight earlier that we are restricting to limited communication scenario. Thus, we do not allow here any error correcting protocols. Because for those protocols also, one needs to apply extra amount of classical communication. 

We have explored explicit cases where the resource state, which cannot lead to zero fidelity deviation in a protocol when the classical communication is noiseless, may lead to zero fidelity deviation when the classical communication is noisy without compromising the non-classical fidelity. This is clearly depicting the fact that in the noisy environment zero fidelity deviation is achievable. We also demonstrate scenarios within the present protocol, where the fidelity deviation increases if the entanglement of the resource state is increased. So, higher amount of entanglement may not always certify the desired quantum teleportation.

For further studies, we have to explore more about different situations where in the presence of noise zero fidelity deviation is achievable. Furthermore, it is also important to explore if maximum fidelity can be achieved along with zero fidelity deviation.

\section*{Acknowledgments}
P.B. acknowledges Prof. Somshubhro Bandyopadhyay for helpful discussions. R.S. acknowledges financial supports from SERB MATRICS MTR/2017/000431 and DST/ICPS/QuST/ Theme-2/2019/General Project No.~Q-90.

\bibliographystyle{elsarticle-num}
\bibliography{ref}
\end{document}